\documentclass[preprint,letterpaper]{aastex}

\usepackage{amsmath, amssymb, graphicx, color}
\usepackage{natbib}
\allowdisplaybreaks

\def\sss{\scriptscriptstyle}

\begin{document}

\title{Roche Lobes in the Second Post-Newtonian Approximation}
\author{Sa{\v s}a Ratkovi{\' c}}
\email{ratkovic@grad.physics.sunysb.edu}
\affil{Department of Physics \& Astronomy, \\
        State University of New York at Stony Brook, \\
        Stony Brook, NY 11794-3800, USA}
\author{Madappa Prakash}
\email{prakash@helios.phy.ohiou.edu}
\affil{Department of Physics \& Astronomy,\\
        Ohio University, \\
	Athens, OH 45701, USA}
\author{James M. Lattimer}
\email{lattimer@mail.astro.sunysb.edu}
\affil{Department of Physics \& Astronomy, \\
        State University of New York at Stony Brook, \\
        Stony Brook, NY 11794-3800, USA}
\shorttitle{2PN Roche Lobes}
\shortauthors{Ratkovi{\' c}, Prakash, and Lattimer}

\begin{abstract}
Close binary systems of compact stars, due to the emission of gravitational
radiation, may evolve into a phase in which the less massive star transfers
mass to its companion. We describe 
mass transfer by using the
model of Roche lobe overflow, in which mass is transferred through the first,
or innermost, Lagrange point. Under conditions in which gravity is strong, the
shapes of the equipotential surfaces and the Roche lobes are modified compared
to the Newtonian case. We present calculations of the Roche lobe utilizing the
second order post-Newtonian (2PN) approximation in the Arnowitt-Deser-Misner
gauge. Heretofore, calculations of the Roche lobe geometry beyond the
Newtonian case have not been available. Beginning from the general N-body
Lagrangian derived by Damour and Sch\"affer, we develop the Lagrangian for a
test particle in the vicinity of two massive compact objects. As an exact
result for the transverse-traceless part of the Lagrangian is not available,
we devise an approximation that is valid 
for regions
close to the less massive star.
We calculate the Roche lobe volumes, and provide a
simple 
fitting formula for the effective Roche lobe radius
analogous to that for the Newtonian case furnished by Eggleton. In contrast to
the Newtonian case, in which the effective Roche radius depends only upon the
mass ratio $q=m_1/m_2$, in the 2PN case the effective Roche lobe radius also
depends on the ratio $z=2 (m_1+m_2)/a$ of the total mass and the 
orbital separation.
\end{abstract}

\keywords{
relativity
---
binaries: close
---
stars: mass loss
}

\newpage


\section{Introduction}

During the evolution of a close binary system involving compact stars, the
stellar separation shrinks due to the emission of gravitational waves. In the
event that the stars are not of equal mass, and the less massive star has a
larger radius than its companion, mass transfer may ultimately occur. Gravity
wave emission generally causes the mutual orbit to circularize
\citep{PETERS64}. For circular orbits, conservative mass transfer can be
modelled as Roche lobe overflow under the assumption
that the star is not significantly disrupted due to tidal interactions.
The Roche lobe is the innermost gravitational plus centrifugal
equipotential surface encompassing both stars.

In the model Roche lobe overflow, the radius of the less massive star
is compared to the effective radius of its Roche lobe. Once the two
radii become equal, because the Roche lobe radius decreases due to
orbital decay, the star fills its Roche lobe and mass transfer occurs
through the first, or innermost, Lagrange point $L_1$. Lying on the
Roche lobe, $L_1$ is located between the two stars on the axis
connecting their centers and is also a saddle point of the
gravitational plus centrifugal potential between the two stars. Due to
its saddle point nature, the first Lagrange point acts as a
gravitational funnel through which mass transfer occurs.

Values of the Roche lobe radii as a function of orbital separation and
mass ratio $q=m_1/m_2$, where $m_1$ refers to the lighter star, have been
tabulated by \citet{KOPAL1} for the Newtonian case. \citet{PACZYNSKI1} and
\citet{EGGLETON1} have given analytical  fits. We use Eggleton's functional
form, which has the advantage of being a continuous function of $q$, as a
template in our work.

In this work, we carry out calculations of Roche lobes beyond the
Newtonian case. We employ the Arnowitt-Deser-Misner (ADM) form of
post-Newtonian expansion and use the corresponding Lagrangian at the
second order (2PN) level wherein terms up to $(M/r)^2$, where
$M=m_1+m_2$ and $r$ is the distance, are retained.  The
same procedure as used in the Newtonian case for finding
the Roche lobes is utilized.  Our strategy is to
(i) construct the effective potential for the point particle in the
vicinity of two stars (the 3--body problem) in the co--rotating frame;
(ii) evaluate equipotential surfaces and calculate the corresponding
effective Roche volume and radius for this potential; and
(iii) provide new fitting formulae as Eggleton did for applications involving
mass transfer.

The organization of this work is as follows. In \S \ref{SEC:2PNPOTENTIAL}, we
calculate the effective potential for three bodies at the 2PN level. We
establish the Lagrangian in \S \ref{SEC:L2PN}. The transverse-traceless part
of the Lagrangian is evaluated explictly in \S\ref{SEC:UTT} through the
introduction of an approximation valid for regions near $m_1$ for test
particles. In \S \ref{SEC:2PNROCHE}, we evaluate the Roche lobes and their
effective radii as a function of $q$ and a relativity parameter for this
potential, and provide a simple analytical fit. In this section, we also show
the impact of post--Newtonian corrections on the positions of the Lagrange
points and on the position of the center of mass. Our conclusions are
contained in \S \ref{sec:conclusion}.

\begin{figure}[!ht]
\begin{center}
\includegraphics[width=0.9\textwidth]{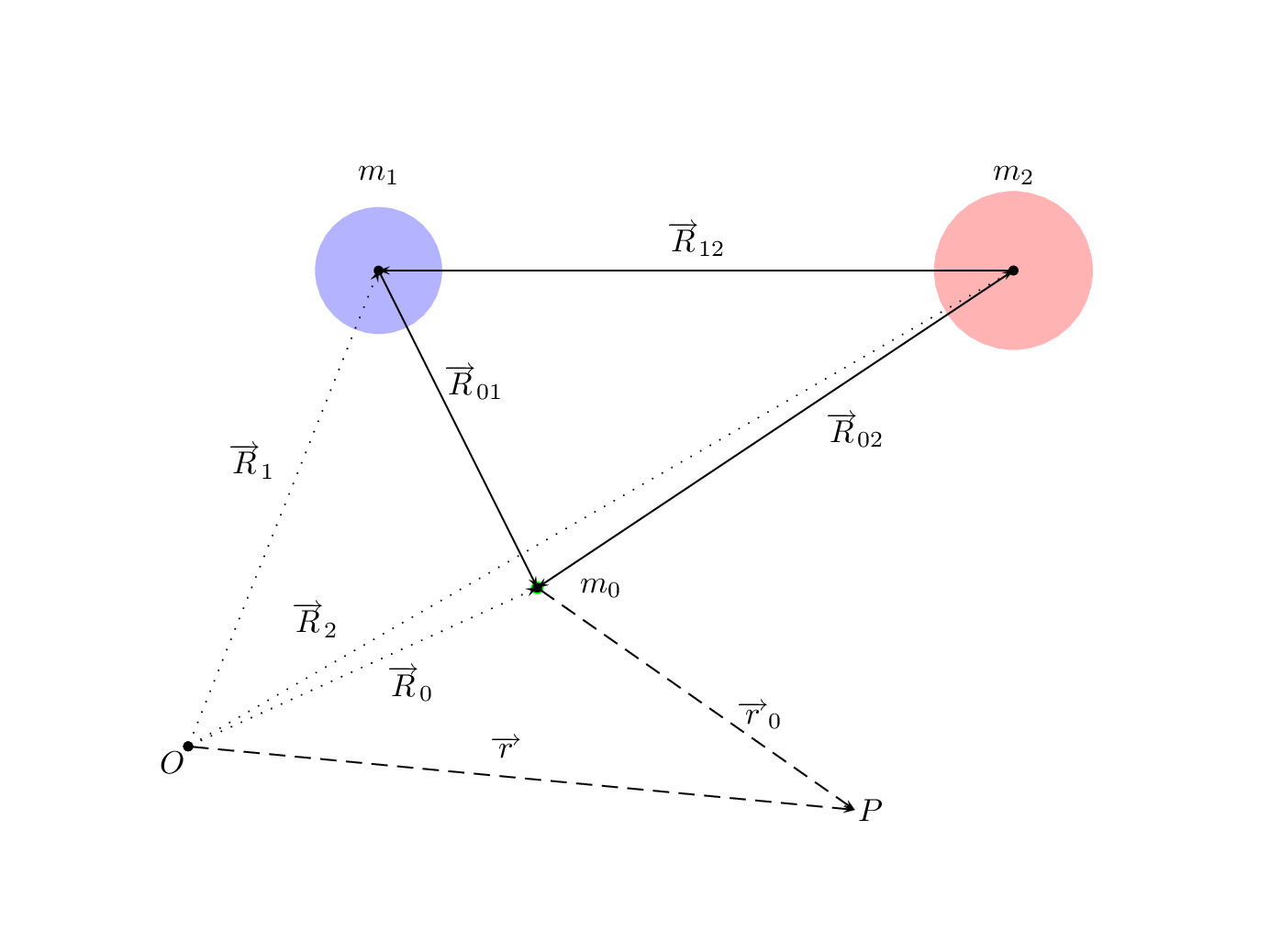}
\caption{
%
The notation used in the
evaluation of the Roche lobes in the 2PN approximation.  Stellar
masses are denoted by $m_1$ and $m_2$ and the point-particle mass is
taken to be $m_0$. Vectors ${\bf R}_A$ ($A=0,1,2$) denote positions
of the three bodies with respect to the origin $O$, ${\bf r}$ is the
position of a generic point $P$, and ${\bf r}_A$ is the position of
this point with respect to the mass $m_A$ (we show only ${\bf
r}_0$). 
The vectors ${\bf R}_{AB}$ indicate positions of the three bodies with
respect to each other.
\label{FIG:3BODY}}
\end{center}
\end{figure}

\section{The 2PN potential for 3 bodies}
\label{SEC:2PNPOTENTIAL}

Roche lobes are defined through the acceleration that a point-like
particle feels in the frame that is co-rotating with the two massive
objects. While velocities of all three bodies disappear in this frame,
accelerations are vanishing only for the two massive objects with
masses $m_1$ and $m_2$. The effective potential causes acceleration on
the third object, the point-particle with mass $m_0$. Starting from
the N-body Lagrangian in ADM coordinates derived by \citet{DAMOUR85},
we obtain the 3--body Lagrangian for the situation depicted in Figure
\ref{FIG:3BODY}. 
We adopt the convention
in which we denote masses with uppercase Latin indices ($A,B,\ldots$)
and coordinates with lowercase Latin indices ($i,j,\ldots$).  Also, we
use units such that $G=1$ and $c=1$. We express all inertial-frame
velocities in terms of the rotating-frame velocities and the remaining
rotationally induced part:
\begin{eqnarray}
{\bf v}_A = {\bf v}_A^{rot} + {\bf \omega}\times{\bf r}_A\, , 
\end{eqnarray}
where ${\bf \omega}$ is the angular frequency of the rotating frame.
Setting ${\bf v}_A^{rot}=0$ for $A=\{0,1,2\}$, and $\dot{\bf
v}_A^{rot}=0$ for $A=\{1,2\}$, but keeping $\dot{\bf v}_0^{rot}\neq
0$ enables us to find the acceleration on the point--like particle
(body $0$) from
\begin{eqnarray}
   m_0\ddot{\xi}_i = \frac{d}{dt}\Big( \frac{\partial L}{\partial
   \dot{\xi}_i}\Big)\Bigg|_{v_A^{c.r.}=0}\, ,
\end{eqnarray}
where we have denoted the coordinate of the point--like body in the
co--rotating frame by ${\bf \xi}$. We have
\begin{eqnarray}
  {\bf \xi_i} = 0\, ,&&\qquad  \dot{\bf \xi} = 0\, ,\qquad \textrm{and}\qquad 
  \ddot{\bf \xi_i} \ne 0\, ,
\end{eqnarray}
since $\dot{\bf v}_0^{c.r.} = \ddot{\bf \xi}$.  We can find the
effective potential for the particle $0$ by separating 
out
the ``kinetic''
part of the Lagrangian that contains terms that are quadratic in
$\dot{\bf \xi}$ and by treating the remaining part of the Lagrangian
as the effective potential that we have to determine. It is
straighforward to verify that this approach yields the Euler-Lagrange
equations for $\xi_i$. After setting $\dot{\bf \xi}=0$, we use the
resulting potential in order to trace the equipotential surfaces that
correspond to the Roche lobes.


\subsection{The 2PN Lagrangian}
\label{SEC:L2PN}

The computation of the effective Roche radii requires the effective
potential that acts on a point particle in the vicinity of two massive
bodies. In order to improve upon the existing Newtonian results, we
utilize results that were obtained by using the post-Newtonian
approximation of general relativity.

The Roche problem requires a three--body Lagrangian in the case in which
one of the bodies is a point-like particle of infinitesimal mass. Such
results were derived for the more general N-body case by
\cite{DAMOUR85} who retained terms up to the second order (2PN) in
$M/r$ in the Arnowitt--Deser--Misner (ADM) coordinate gauge.

For completeness, we list the main results of \citet{DAMOUR85}
here. The Newtonian, or zeroth order, result is familiar:
\begin{eqnarray}
L_N
&=&
\frac{1}{2}\sum_A m_A v_A^2 + 
\frac{1}{2}\sum_{A, B\ne A} \frac{m_A m_B}{r_{AB}}\, ,
\end{eqnarray}
where $r_{AB}\equiv |{\bf R}_{AB}|$. The first order post-Newtonian
correction is
\begin{eqnarray}
L_2
&=&
\frac{1}{8}\sum_A m_A v_A^4 
\nonumber \\
&&
+
\frac{1}{4}\sum_{A, B\ne A} \frac{m_A m_B}{r_{AB}}
\Big\{ 
  6 v_A^2 - 7 \left( {\bf v}_A \cdot {\bf v}_B \right)
  - \left( {\bf n}_{AB} \cdot {\bf v}_A \right)
  \left( {\bf n}_{AB} \cdot {\bf v}_B \right)
\Big\}
\nonumber \\
&&- 
\frac{1}{2}\sum_{A \ne B\ne C} \frac{m_A m_B m_C}{r_{AB}
  r_{AC}}\, ,
\label{EQ:NBODYL2}
\end{eqnarray}
whereas the second order post-Newtonian contribution is
\begin{eqnarray}
\label{EQ:NBODYL4}
L_4
&=&
\frac{1}{16} \sum_{A} m_A v_A^6 
+
\frac{3}{8} \sum_{
%
\genfrac{}{}{0pt}{}{A, B\ne A,}{C\ne B, D\ne C}
}  
\frac{m_A m_B m_C m_D}{r_{AB}\, r_{BC}\, r_{CD}}
+
\frac{1}{4} \sum_{
%
\genfrac{}{}{0pt}{}{A, B\ne A,}{C\ne A, D\ne A}
}  
\frac{m_A m_B m_C m_D}{r_{AB}\, r_{AC}\, r_{AD}}
-
U_{TT}
\nonumber \\
&& 
+
\frac{1}{4} \sum_{
\genfrac{}{}{0pt}{}{A, B\ne A,}{C\ne A}
}  
\frac{m_A m_B m_C}{r_{AB}\, r_{AC}}
\Big\{
  9 v_A^2 - 7v_B^2 - 17\left( {\bf v}_A \cdot {\bf v}_B \right)
  + \left( {\bf n}_{AB}\cdot {\bf v}_{A} \right)
    \left( {\bf n}_{AB}\cdot {\bf v}_{B} \right)
\nonumber \\
&& \qquad 
  + {\left( {\bf n}_{AB}\cdot {\bf v}_{B} \right)}^2
  + 16 \left( {\bf v}_{B}\cdot {\bf v}_{C} \right)
\Big\}
\nonumber \\
&& 
+
\frac{1}{8} \sum_{
\genfrac{}{}{0pt}{}{A, B\ne A,}{C\ne A}
}  
\frac{m_A m_B m_C}{r_{AB}^2}
\Big\{
  - 5 \left( {\bf n}_{AB}\cdot {\bf n}_{AC} \right) v_C^2
  + \left( {\bf n}_{AB}\cdot {\bf n}_{AC} \right)
    { \left( {\bf n}_{AC}\cdot {\bf v}_{C} \right) }^2
\nonumber \\
&& \qquad
  - 2 \left( {\bf n}_{AB}\cdot {\bf v}_{A} \right)
    \left( {\bf n}_{AC}\cdot {\bf v}_{C} \right)
  - 2 \left( {\bf n}_{AB}\cdot {\bf v}_{B} \right)
    \left( {\bf n}_{AC}\cdot {\bf v}_{C} \right)
  + 14 \left( {\bf n}_{AB}\cdot {\bf v}_{C} \right)
    \left( {\bf n}_{AC}\cdot {\bf v}_{C} \right)
\Big\}
\nonumber \\
&& 
-
\frac{1}{2} \sum_{
\genfrac{}{}{0pt}{}{A, B\ne A,}{C\ne A,B}
}  
\frac{m_A m_B m_C}{{\left(r_{AB} + r_{BC} + r_{CA}\right)}^2}
\Big\{
  3 \left( {\bf n}_{AB} + {\bf n}_{AC} \right) \cdot {\bf v}_A
    \left( {\bf n}_{AB} - {\bf n}_{BC} \right) \cdot {\bf v}_B
\nonumber \\
&& \qquad
  +
  \left( {\bf n}_{AB} + {\bf n}_{AC} \right) \cdot {\bf v}_A
    \left( {\bf n}_{AB} - {\bf n}_{BC} \right) \cdot {\bf v}_A
  +
  8 \left( {\bf n}_{AB} + {\bf n}_{AC} \right) \cdot {\bf v}_A
    \left( {\bf n}_{AB} - {\bf n}_{BC} \right) \cdot {\bf v}_C
\nonumber \\
&& \qquad
  -
  16 \left( {\bf n}_{AB} + {\bf n}_{AC} \right) \cdot {\bf v}_C
    \left( {\bf n}_{AB} - {\bf n}_{BC} \right) \cdot {\bf v}_A
  +
  4 \left( {\bf n}_{AB} + {\bf n}_{AC} \right) \cdot {\bf v}_C
    \left( {\bf n}_{AB} - {\bf n}_{BC} \right) \cdot {\bf v}_C
\Big\}
\nonumber \\
&& 
+
\frac{1}{2} \sum_{
\genfrac{}{}{0pt}{}{A, B\ne A,}{C\ne A,B}
}  
\frac{m_A m_B m_C}{\left(r_{AB} + r_{BC} + r_{CA}\right) r_{AB}}
\Bigg\{
  3 \Big[
    \left( {\bf v}_{A} \cdot {\bf v}_{B} \right) 
    -
    \left( {\bf n}_{AB} \cdot {\bf v}_A \right)
    \left( {\bf n}_{AB} \cdot {\bf v}_B \right)
    \Big]
\nonumber \\
&& \qquad
  +
  \Big[
    v_A^2 - {\left( {\bf n}_{AB} \cdot {\bf v}_A \right)}^2
    \Big]
  -
  8 \Big[
    \left( {\bf v}_{A} \cdot {\bf v}_{C} \right) 
    -
    \left( {\bf n}_{AB} \cdot {\bf v}_A \right)
    \left( {\bf n}_{AB} \cdot {\bf v}_C \right)
    \Big]
\nonumber \\
&& \qquad
  +
  4 \Big[
    v_C^2 - {\left( {\bf n}_{AB} \cdot {\bf v}_C \right)}^2
    \Big]
\Bigg\}
\nonumber \\
&& 
-
\frac{1}{4} \sum_{A, B\ne A}
\frac{m_A m_B^2}{r_{AB}^2}
\Big\{
  v_A^2 + v_B^2 - 2 \left( {\bf v}_A \cdot {\bf v}_B \right) 
\Big\}
\nonumber \\
&& 
+
\frac{1}{16} \sum_{A, B\ne A}  
\frac{m_A m_B}{r_{AB}}
\Big\{
  14 v_A^4 
  - 28 v_A^2\left( {\bf v}_A \cdot {\bf v}_B \right)
  - 4 v_A^2   \left( {\bf n}_{AB} \cdot {\bf v}_{A} \right) 
    \left( {\bf n}_{AB} \cdot {\bf v}_{B} \right)
  + 11 v_A^2 v_B^2 
\nonumber \\
&& \qquad
  + 2 {\left( {\bf v}_A \cdot {\bf v}_A \right)}^2
  - 10 v_A^2 {\left( {\bf n}_{AB} \cdot {\bf v}_B \right)}^2
  + 12 \left( {\bf v}_A \cdot {\bf v}_B \right)
    \left( {\bf n}_{AB} \cdot {\bf v}_A \right)
    \left( {\bf n}_{AB} \cdot {\bf v}_B \right)
\nonumber \\
&& \qquad
  + 3 { \left( {\bf n}_{AB} \cdot {\bf v}_A \right) }^2
    { \left( {\bf n}_{AB} \cdot {\bf v}_B \right) }^2
\Big\}\, .
\end{eqnarray}
The term $U_{TT}$ refers to the transverse-traceless part of the Lagrangian
potential which requires special treatment and is explicitly evaluated in 
\S\ref{SEC:UTT}.

We now specialize to the case of three bodies with masses $m_0$, $m_1$, and
$m_2$. For the Newtonian and the first order post-Newtonian correction, we
obtain
\begin{eqnarray}
L_N
&=&
\frac{1}{2}
\Big\{
m_0 v_0^2 + m_1 v_1^2 + m_2 v_2^2
\Big\}
+
\frac{m_0 m_1}{r_{01}}
+
\frac{m_0 m_2}{r_{02}}
+
\frac{m_1 m_2}{r_{12}}\, ,
\end{eqnarray}
and
\begin{eqnarray}
L_2
&=&
\frac{1}{8}
\Big\{ 
m_0 v_0^4 + m_1 v_1^4 + m_2 v_2^4
\Big\}
\nonumber \\
&&
+ 
\frac{1}{4}
\Bigg\{ 
\frac{m_0 m_1}{r_{01}}
\Big[
  6 v_0^2  + 6 v_1^2 - 14 \left( {\bf v}_0 \cdot {\bf v}_1 \right)
  - 2 \left( {\bf n}_{01} \cdot {\bf v}_0 \right)
    \left( {\bf n}_{01} \cdot {\bf v}_1 \right)
\Big]
\nonumber \\
&& \hspace{20pt}
+
\frac{m_0 m_2}{r_{02}}
\Big[
  6 v_0^2  + 6 v_2^2 - 14 \left( {\bf v}_0 \cdot {\bf v}_2 \right)
  - 2 \left( {\bf n}_{02} \cdot {\bf v}_0 \right)
    \left( {\bf n}_{02} \cdot {\bf v}_2 \right)
\Big]
\nonumber \\
&& \hspace{20pt}
+
\frac{m_1 m_2}{r_{12}}
\Big[
  6 v_1^2  + 6 v_2^2 - 14 \left( {\bf v}_1 \cdot {\bf v}_2 \right)
  - 2 \left( {\bf n}_{12} \cdot {\bf v}_1 \right)
    \left( {\bf n}_{12} \cdot {\bf v}_2 \right)
\Big]
\Bigg\}
\nonumber \\
&&- 
\frac{m_0 m_1 m_2}{r_{01} r_{02} r_{12}}
\Big\{
  r_{01} + r_{02} + r_{12}
\Big\}\, .
\end{eqnarray}

Because the second order post-Newtonian correction $L_4$ is small, and we
are interested in computing the equipotential surfaces for a test particle, we
assume
\begin{eqnarray}
\label{EQ:APPROXM}
m_0 \ll m_1\, ,m_2.
\end{eqnarray}
We decompose the second order correction in order to facilitate its
presentation:
%
%
\begin{eqnarray}
L_4
&=& 
\sum_{a=i}^{xi} L_4^{(a)} + \mathcal{O}\left(m_0^2\right)
\end{eqnarray}
and we drop terms of $\mathcal{O}\left(m_0^2\right)$. The decomposition is
evident by comparing equation (\ref{EQ:NBODYL4}) with the following:
\begin{eqnarray}
L_4^{(i)} 
&=&
\frac{1}{16} 
\Big\{
m_0 v_0^6  + m_1 v_1^6 + m_2 v_2^6 
\Big\}\, ,
\label{EQ:2PNL01}
\end{eqnarray}
\begin{eqnarray}
L_4^{(ii)}
&=&
\frac{3}{4} 
\Big\{
\frac{m_0 m_1^2 m_2}{r_{01} r_{12}^2}
+
\frac{m_0 m_1 m_2^2}{r_{02} r_{12}^2}
+
\frac{m_0 m_1^2 m_2}{r_{01}^2 r_{12}}
+
\frac{m_0 m_1 m_2^2}{r_{02}^2 r_{12}}
+
\frac{m_0 m_1^2 m_2}{r_{01} r_{12} r_{02}}
+
\frac{m_0 m_1 m_2^2}{r_{01} r_{12} r_{02}}
\nonumber \\ 
&&
\qquad
+
\frac{m_1^2 m_2^2}{r_{12}^3}
+
\frac{m_0 m_1^2 m_2}{r_{01}^2 r_{02}}
+
\frac{m_0 m_1 m_2^2}{r_{01} r_{02}^2}
+
\frac{m_0 m_1^2 m_2}{r_{02} r_{12}^2}
+
\frac{m_0 m_1 m_2^2}{r_{01} r_{12}^2}
\Big\}\, ,
\label{EQ:2PNL02}
\end{eqnarray}
\begin{eqnarray}
L_4^{(iii)}
&=&
\frac{1}{4} 
\Big\{
\frac{m_0 m_1^3}{r_{01}^3}
+
\frac{m_0 m_2^3}{r_{02}^3}
+
\frac{m_1 m_2^3}{r_{12}^3}
+
\frac{m_1^3 m_2}{r_{12}^3}
\Big\}\, ,
\label{EQ:2PNL03}
\end{eqnarray}
\begin{eqnarray}
L_4^{(iv)}
&=&
-
U_{TT}\, , 
\label{EQ:2PNL04}
\end{eqnarray}
\begin{eqnarray}
L_4^{(v)}
&=&
\frac{m_0 m_1^2}{r_{01}^2}
\Big\{
  \frac{9}{4}\, v_0^2
  + \frac{13}{8}\, v_1^2
  - \frac{17}{4}\, \left( {\bf v}_0 \cdot {\bf v}_1 \right)
  + \frac{15}{8}\, {\left( {\bf n}_{01} \cdot {\bf v}_1 \right)}^2
\Big\}
\nonumber \\
&& 
+
\frac{m_0 m_2^2}{r_{02}^2}
\Big\{
  \frac{9}{4}\, v_0^2
  + \frac{13}{8}\, v_2^2
  - \frac{17}{4}\, \left( {\bf v}_0 \cdot {\bf v}_2 \right)
  + \frac{15}{8}\, {\left( {\bf n}_{02} \cdot {\bf v}_2 \right)}^2
\Big\}
\nonumber \\
&& 
+
\frac{m_1 m_2^2}{r_{12}^2}
\Big\{
  \frac{9}{4}\, v_1^2
  + \frac{13}{8}\, v_2^2
  - \frac{17}{4}\, \left( {\bf v}_1 \cdot {\bf v}_2 \right)
  + \frac{15}{8}\, {\left( {\bf n}_{12} \cdot {\bf v}_2 \right)}^2
\Big\}
\nonumber \\
&& 
+
\frac{m_1^2 m_2}{r_{12}^2}
\Big\{
  \frac{9}{4}\, v_2^2
  + \frac{13}{8}\, v_1^2
  - \frac{17}{4}\, \left( {\bf v}_1 \cdot {\bf v}_2 \right)
  + \frac{15}{8}\, {\left( {\bf n}_{12} \cdot {\bf v}_1 \right)}^2
\Big\}\, ,
\nonumber \\ 
\label{EQ:2PNL05}
\end{eqnarray}
\begin{eqnarray}
L_4^{(vi)}
&=&
\frac{m_0 m_1 m_2}{4\, r_{01} r_{12}}
\Big\{
  18 v_1^2
  - 7 v_0^2
  - 17 \left( {\bf v}_0 \cdot {\bf v}_1 \right)
  + \left( {\bf n}_{01} \cdot {\bf v}_0 \right)
    \left( {\bf n}_{01} \cdot {\bf v}_1 \right)
  + {\left( {\bf n}_{01} \cdot {\bf v}_0 \right)}^2
\nonumber \\ 
&& \qquad
  + 32 \left( {\bf v}_0 \cdot {\bf v}_2 \right) 
  - 7 v_2^2
  - 17 \left( {\bf v}_1 \cdot {\bf v}_2 \right)
  + \left( {\bf n}_{12} \cdot {\bf v}_1 \right)
    \left( {\bf n}_{12} \cdot {\bf v}_2 \right)
  + {\left( {\bf n}_{12} \cdot {\bf v}_2 \right)}^2
\Big\}
\nonumber \\
&&
+
\frac{m_0 m_1 m_2}{4\, r_{01} r_{02}}
\Big\{
  18 v_0^2
  - 7 v_1^2
  - 17 \left( {\bf v}_0 \cdot {\bf v}_1 \right)
  + \left( {\bf n}_{01} \cdot {\bf v}_0 \right)
    \left( {\bf n}_{01} \cdot {\bf v}_1 \right)
  + {\left( {\bf n}_{01} \cdot {\bf v}_1 \right)}^2
\nonumber \\ 
&& \qquad
  + 32 \left( {\bf v}_1 \cdot {\bf v}_2 \right) 
  - 7 v_2^2
  - 17 \left( {\bf v}_0 \cdot {\bf v}_2 \right)
  + \left( {\bf n}_{02} \cdot {\bf v}_0 \right)
    \left( {\bf n}_{02} \cdot {\bf v}_2 \right)
  + {\left( {\bf n}_{02} \cdot {\bf v}_2 \right)}^2
\Big\}
\nonumber \\
&&
+
\frac{m_0 m_1 m_2}{4\, r_{02} r_{12}}
\Big\{
  18 v_2^2
  - 7 v_0^2
  - 17 \left( {\bf v}_0 \cdot {\bf v}_2 \right)
  + \left( {\bf n}_{02} \cdot {\bf v}_0 \right)
    \left( {\bf n}_{02} \cdot {\bf v}_2 \right)
  + {\left( {\bf n}_{02} \cdot {\bf v}_0 \right)}^2
\nonumber \\ 
&& \qquad
  + 32 \left( {\bf v}_0 \cdot {\bf v}_1 \right) 
  - 7 v_1^2
  - 17 \left( {\bf v}_1 \cdot {\bf v}_2 \right)
  + \left( {\bf n}_{12} \cdot {\bf v}_1 \right)
    \left( {\bf n}_{12} \cdot {\bf v}_2 \right)
  + {\left( {\bf n}_{12} \cdot {\bf v}_1 \right)}^2
\Big\}\, ,
\label{EQ:2PNL06}
\end{eqnarray}
\begin{eqnarray}
L_4^{(vii)}
&=&
\frac{m_0 m_1 m_2}{8\, r_{01}^2}
\Big\{
  - 5 \left( {\bf n}_{01} \cdot {\bf n}_{02} \right) v_2^2
  + \left( {\bf n}_{01} \cdot {\bf n}_{02} \right) 
    {\left( {\bf n}_{02} \cdot {\bf v}_2 \right) }^2
  - 2 \left( {\bf n}_{01} \cdot {\bf v}_0 \right)   
    \left( {\bf n}_{02} \cdot {\bf v}_2 \right)   
\nonumber \\ 
&& \qquad
  - 2 \left( {\bf n}_{01} \cdot {\bf v}_1 \right)   
    \left( {\bf n}_{02} \cdot {\bf v}_2 \right)
  + 14 \left( {\bf n}_{01} \cdot {\bf v}_2 \right)   
    \left( {\bf n}_{02} \cdot {\bf v}_2 \right)   
  + 5 \left( {\bf n}_{01} \cdot {\bf n}_{12} \right) v_2^2
\nonumber \\ 
&& \qquad
  - \left( {\bf n}_{01} \cdot {\bf n}_{12} \right) 
    {\left( {\bf n}_{12} \cdot {\bf v}_2 \right) }^2
  + 2 \left( {\bf n}_{01} \cdot {\bf v}_1 \right)   
    \left( {\bf n}_{12} \cdot {\bf v}_2 \right)   
  + 2 \left( {\bf n}_{01} \cdot {\bf v}_0 \right)   
    \left( {\bf n}_{12} \cdot {\bf v}_2 \right)
\nonumber \\ 
&& \qquad
  - 14 \left( {\bf n}_{01} \cdot {\bf v}_2 \right)   
    \left( {\bf n}_{12} \cdot {\bf v}_2 \right)   
\Big\}
\nonumber \\
&&
+
\frac{m_0 m_1 m_2}{8\, r_{02}^2}
\Big\{
  - 5 \left( {\bf n}_{01} \cdot {\bf n}_{02} \right) v_1^2
  + \left( {\bf n}_{01} \cdot {\bf n}_{02} \right) 
    {\left( {\bf n}_{01} \cdot {\bf v}_1 \right) }^2
  - 2 \left( {\bf n}_{01} \cdot {\bf v}_1 \right)   
    \left( {\bf n}_{02} \cdot {\bf v}_0 \right)   
\nonumber \\ 
&& \qquad
  - 2 \left( {\bf n}_{02} \cdot {\bf v}_2 \right)   
    \left( {\bf n}_{01} \cdot {\bf v}_1 \right)
  + 14 \left( {\bf n}_{01} \cdot {\bf v}_1 \right)   
    \left( {\bf n}_{02} \cdot {\bf v}_1 \right)   
  - 5 \left( {\bf n}_{02} \cdot {\bf n}_{12} \right) v_1^2
\nonumber \\ 
&& \qquad
  + \left( {\bf n}_{02} \cdot {\bf n}_{12} \right) 
    {\left( {\bf n}_{12} \cdot {\bf v}_1 \right) }^2
  - 2 \left( {\bf n}_{02} \cdot {\bf v}_2 \right)   
    \left( {\bf n}_{12} \cdot {\bf v}_1 \right)   
  - 2 \left( {\bf n}_{02} \cdot {\bf v}_0 \right)   
    \left( {\bf n}_{12} \cdot {\bf v}_1 \right)
\nonumber \\ 
&& \qquad
  + 14 \left( {\bf n}_{02} \cdot {\bf v}_1 \right)   
    \left( {\bf n}_{12} \cdot {\bf v}_1 \right)   
\Big\}
\nonumber \\
&&
+
\frac{m_0 m_1 m_2}{8\, r_{12}^2}
\Big\{
  5 \left( {\bf n}_{01} \cdot {\bf n}_{12} \right) v_0^2
  - \left( {\bf n}_{01} \cdot {\bf n}_{12} \right) 
    {\left( {\bf n}_{01} \cdot {\bf v}_0 \right) }^2
  + 2 \left( {\bf n}_{12} \cdot {\bf v}_1 \right)   
    \left( {\bf n}_{01} \cdot {\bf v}_0 \right)   
\nonumber \\ 
&& \qquad
  + 2 \left( {\bf n}_{12} \cdot {\bf v}_2 \right)   
    \left( {\bf n}_{01} \cdot {\bf v}_0 \right)
  - 14 \left( {\bf n}_{12} \cdot {\bf v}_0 \right)   
    \left( {\bf n}_{01} \cdot {\bf v}_0 \right)   
  - 5 \left( {\bf n}_{12} \cdot {\bf n}_{02} \right) v_0^2
\nonumber \\ 
&& \qquad
  + \left( {\bf n}_{12} \cdot {\bf n}_{02} \right) 
    {\left( {\bf n}_{02} \cdot {\bf v}_0 \right) }^2
  - 2 \left( {\bf n}_{12} \cdot {\bf v}_2 \right)   
    \left( {\bf n}_{02} \cdot {\bf v}_0 \right)   
  - 2 \left( {\bf n}_{12} \cdot {\bf v}_1 \right)   
    \left( {\bf n}_{02} \cdot {\bf v}_0 \right)
\nonumber \\ 
&& \qquad
  + 14 \left( {\bf n}_{12} \cdot {\bf v}_0 \right)   
    \left( {\bf n}_{02} \cdot {\bf v}_0 \right)   
\Big\}\, ,
\label{EQ:2PNL07}
\end{eqnarray}
\begin{eqnarray}
L_4^{(viii)}
&=&
-
\frac{1}{2} \, 
\frac{m_0 m_1 m_2}{{\left( r_{01} + r_{12} + r_{02} \right)}^2} 
\Bigg\{
  {\left( {\bf n}_{01} \cdot {\bf v}_0 \right) }^2
  + {\left( {\bf n}_{01} \cdot {\bf v}_1 \right) }^2
  + {\left( {\bf n}_{12} \cdot {\bf v}_1 \right) }^2
  + {\left( {\bf n}_{12} \cdot {\bf v}_2 \right) }^2
\nonumber \\ 
&& \qquad
  + {\left( {\bf n}_{02} \cdot {\bf v}_0 \right) }^2
  + {\left( {\bf n}_{02} \cdot {\bf v}_2 \right) }^2
+ 
8 \Big[
  {\left( {\bf n}_{01} \cdot {\bf v}_2 \right) }^2
  + {\left( {\bf n}_{02} \cdot {\bf v}_1 \right) }^2
  + {\left( {\bf n}_{12} \cdot {\bf v}_0 \right) }^2
  \Big]
\nonumber \\ 
&& \qquad
+ 
32 \Big[
  \left( {\bf n}_{01} \cdot {\bf v}_2 \right)
    \left( {\bf n}_{12} \cdot {\bf v}_0 \right)
  - \left( {\bf n}_{02} \cdot {\bf v}_1 \right)
    \left( {\bf n}_{12} \cdot {\bf v}_0 \right)
  - \left( {\bf n}_{01} \cdot {\bf v}_2 \right)
    \left( {\bf n}_{02} \cdot {\bf v}_1 \right)
  \Big]
\nonumber \\ 
&& \qquad
+ 
10 \Big[
  \left( {\bf n}_{02} \cdot {\bf v}_0 \right)
    \left( {\bf n}_{12} \cdot {\bf v}_1 \right)
  - \left( {\bf n}_{01} \cdot {\bf v}_0 \right)
    \left( {\bf n}_{12} \cdot {\bf v}_2 \right)
  - \left( {\bf n}_{01} \cdot {\bf v}_1 \right)
    \left( {\bf n}_{02} \cdot {\bf v}_2 \right)
  \Big]
\nonumber \\ 
&& \qquad
+
6 \Big[
  \left( {\bf n}_{01} \cdot {\bf v}_0 \right)
    \left( {\bf n}_{01} \cdot {\bf v}_1 \right)
  + \left( {\bf n}_{02} \cdot {\bf v}_0 \right)
    \left( {\bf n}_{02} \cdot {\bf v}_2 \right)
  + \left( {\bf n}_{12} \cdot {\bf v}_1 \right)
    \left( {\bf n}_{12} \cdot {\bf v}_2 \right)
\nonumber \\ 
&& \qquad
  - \left( {\bf n}_{01} \cdot {\bf v}_0 \right)
    \left( {\bf n}_{02} \cdot {\bf v}_0 \right)
  + \left( {\bf n}_{01} \cdot {\bf v}_1 \right)
    \left( {\bf n}_{12} \cdot {\bf v}_1 \right)
  - \left( {\bf n}_{12} \cdot {\bf v}_2 \right)
    \left( {\bf n}_{02} \cdot {\bf v}_2 \right)
  \Big]
\nonumber \\ 
&& \qquad
+ 
18 \Big[
  \left( {\bf n}_{01} \cdot {\bf v}_0 \right)
    \left( {\bf n}_{12} \cdot {\bf v}_1 \right)
  - \left( {\bf n}_{02} \cdot {\bf v}_0 \right)
    \left( {\bf n}_{01} \cdot {\bf v}_1 \right)
  - \left( {\bf n}_{02} \cdot {\bf v}_0 \right)
    \left( {\bf n}_{12} \cdot {\bf v}_2 \right)
\nonumber \\ 
&& \qquad
  - \left( {\bf n}_{02} \cdot {\bf v}_2 \right)
    \left( {\bf n}_{01} \cdot {\bf v}_0 \right)
  + \left( {\bf n}_{12} \cdot {\bf v}_2 \right)
    \left( {\bf n}_{01} \cdot {\bf v}_1 \right)
  - \left( {\bf n}_{12} \cdot {\bf v}_1 \right)
    \left( {\bf n}_{02} \cdot {\bf v}_2 \right)
  \Big]
\nonumber \\ 
&& \qquad
+ 
8 \Big[
  - \left( {\bf n}_{01} \cdot {\bf v}_0 \right)
    \left( {\bf n}_{12} \cdot {\bf v}_0 \right)
  + \left( {\bf n}_{02} \cdot {\bf v}_0 \right)
    \left( {\bf n}_{12} \cdot {\bf v}_0 \right)
  - \left( {\bf n}_{01} \cdot {\bf v}_2 \right)
    \left( {\bf n}_{12} \cdot {\bf v}_2 \right)
\nonumber \\ 
&& \qquad
  + \left( {\bf n}_{02} \cdot {\bf v}_2 \right)
    \left( {\bf n}_{01} \cdot {\bf v}_2 \right)
  + \left( {\bf n}_{02} \cdot {\bf v}_1 \right)
    \left( {\bf n}_{12} \cdot {\bf v}_1 \right)
  + \left( {\bf n}_{01} \cdot {\bf v}_1 \right)
    \left( {\bf n}_{02} \cdot {\bf v}_1 \right)
  \Big]
\nonumber \\ 
&& \qquad
+ 
8 \Big[
  - \left( {\bf n}_{01} \cdot {\bf v}_2 \right)
    \left( {\bf n}_{01} \cdot {\bf v}_0 \right)
  - \left( {\bf n}_{01} \cdot {\bf v}_2 \right)
    \left( {\bf n}_{01} \cdot {\bf v}_1 \right)
  - \left( {\bf n}_{02} \cdot {\bf v}_1 \right)
    \left( {\bf n}_{02} \cdot {\bf v}_0 \right)
\nonumber \\ 
&& \qquad
  - \left( {\bf n}_{02} \cdot {\bf v}_1 \right)
    \left( {\bf n}_{02} \cdot {\bf v}_2 \right)
  - \left( {\bf n}_{12} \cdot {\bf v}_0 \right)
    \left( {\bf n}_{12} \cdot {\bf v}_1 \right)
  - \left( {\bf n}_{12} \cdot {\bf v}_0 \right)
    \left( {\bf n}_{12} \cdot {\bf v}_2 \right)
  \Big]
\nonumber \\ 
&& \qquad
+ 
24 \Big[
  \left( {\bf n}_{01} \cdot {\bf v}_2 \right)
    \left( {\bf n}_{02} \cdot {\bf v}_0 \right)
  + \left( {\bf n}_{01} \cdot {\bf v}_0 \right)
    \left( {\bf n}_{02} \cdot {\bf v}_1 \right)
  + \left( {\bf n}_{02} \cdot {\bf v}_2 \right)
    \left( {\bf n}_{12} \cdot {\bf v}_0 \right)
\nonumber \\ 
&& \qquad
  + \left( {\bf n}_{02} \cdot {\bf v}_1 \right)
    \left( {\bf n}_{12} \cdot {\bf v}_2 \right)
  - \left( {\bf n}_{01} \cdot {\bf v}_2 \right)
    \left( {\bf n}_{12} \cdot {\bf v}_1 \right)
  - \left( {\bf n}_{01} \cdot {\bf v}_1 \right)
    \left( {\bf n}_{12} \cdot {\bf v}_0 \right)
  \Big]
\Bigg\}\, ,
\label{EQ:2PNL08}
\end{eqnarray}
\begin{eqnarray}
L_4^{(ix)}
&=&
\frac{1}{2} \, 
\frac{m_0 m_1 m_2}{\left( r_{01} + r_{12} + r_{02} \right)} 
\Bigg\{
\frac{1}{r_{01}} 
\Big[
  6 \left( {\bf v}_0 \cdot {\bf v}_1 \right) 
  - 6 \left( {\bf n}_{01} \cdot {\bf v}_0 \right)
      \left( {\bf n}_{01} \cdot {\bf v}_1 \right)
  + v_0^2
  - {\left( {\bf n}_{01} \cdot {\bf v}_0 \right) }^2
\nonumber \\ 
&& \qquad
  - 8 \left( {\bf v}_0 \cdot {\bf v}_2 \right)
  + 8 \left( {\bf n}_{01} \cdot {\bf v}_0 \right)
      \left( {\bf n}_{01} \cdot {\bf v}_2 \right)
  + 8 v_2^2
  - 8  {\left( {\bf n}_{01} \cdot {\bf v}_2 \right) }^2
  + v_1^2
  - {\left( {\bf n}_{01} \cdot {\bf v}_1 \right) }^2
\nonumber \\ 
&& \qquad
  - 8 \left( {\bf v}_1 \cdot {\bf v}_2 \right) 
  + 8 \left( {\bf n}_{01} \cdot {\bf v}_1 \right)
      \left( {\bf n}_{01} \cdot {\bf v}_2 \right)
\Big]
\nonumber \\ 
&&
+ 
\frac{1}{r_{12}} 
\Big[
  6 \left( {\bf v}_1 \cdot {\bf v}_2 \right) 
  - 6 \left( {\bf n}_{12} \cdot {\bf v}_1 \right)
      \left( {\bf n}_{12} \cdot {\bf v}_2 \right)
  + v_1^2
  - {\left( {\bf n}_{12} \cdot {\bf v}_1 \right) }^2
  - 8 \left( {\bf v}_0 \cdot {\bf v}_1 \right)
\nonumber \\ 
&& \qquad
  + 8 \left( {\bf n}_{12} \cdot {\bf v}_1 \right)
      \left( {\bf n}_{12} \cdot {\bf v}_0 \right)
  + 8 v_0^2
  - 8  {\left( {\bf n}_{12} \cdot {\bf v}_0 \right) }^2
  + v_2^2
  - {\left( {\bf n}_{12} \cdot {\bf v}_2 \right) }^2
\nonumber \\ 
&& \qquad
  - 8 \left( {\bf v}_0 \cdot {\bf v}_2 \right) 
  + 8 \left( {\bf n}_{12} \cdot {\bf v}_2 \right)
      \left( {\bf n}_{12} \cdot {\bf v}_0 \right)
\Big]
\nonumber \\ 
&&
+ 
\frac{1}{r_{02}} 
\Big[
  6 \left( {\bf v}_0 \cdot {\bf v}_2 \right) 
  - 6 \left( {\bf n}_{02} \cdot {\bf v}_0 \right)
      \left( {\bf n}_{02} \cdot {\bf v}_2 \right)
  + v_0^2
  - {\left( {\bf n}_{02} \cdot {\bf v}_0 \right) }^2
  - 8 \left( {\bf v}_0 \cdot {\bf v}_1 \right)
\nonumber \\ 
&& \qquad
  + 8 \left( {\bf n}_{02} \cdot {\bf v}_0 \right)
      \left( {\bf n}_{01} \cdot {\bf v}_1 \right)
  + 8 v_1^2
  - 8  {\left( {\bf n}_{02} \cdot {\bf v}_1 \right) }^2
  + v_2^2
  - {\left( {\bf n}_{02} \cdot {\bf v}_2 \right) }^2
\nonumber \\ 
&& \qquad
  - 8 \left( {\bf v}_1 \cdot {\bf v}_2 \right) 
  + 8 \left( {\bf n}_{02} \cdot {\bf v}_1 \right)
      \left( {\bf n}_{02} \cdot {\bf v}_2 \right)
\Big]
\Bigg\}\, ,
\label{EQ:2PNL09}
\end{eqnarray}
\begin{eqnarray}
L_4^{(x)}
&=&
-
\frac{m_0 m_1 (m_0 + m_1)}{4 r_{01}^2}
\left(
v_0^2 + v_1^2 
- 2 \left( {\bf v}_0\cdot{\bf v}_1 \right)
\right)
-
\frac{m_0 m_2 (m_0 + m_2)}{4 r_{02}^2}
\left(
v_0^2 + v_2^2 
- 2 \left( {\bf v}_0\cdot{\bf v}_2 \right)
\right)
\nonumber \\
&&
-
\frac{m_1 m_2 (m_1 + m_2)}{4 r_{12}^2}
\left(
v_1^2 + v_2^2 
- 2 \left( {\bf v}_1\cdot{\bf v}_2 \right)
\right)\, ,
\label{EQ:2PNL10}
\end{eqnarray}
\begin{eqnarray}
L_4^{(xi)}
&=&
\frac{m_0 m_1}{16 r_{01}^2}
\Big\{
  14 \left( v_0^4 + v_1^4 \right)
  - 28 \left( {\bf v}_0\cdot{\bf v}_1 \right)
    \left( v_0^2 + v_1^2 \right)
  - 4 \left( {\bf n}_{01} \cdot {\bf v}_0 \right)
    \left( {\bf n}_{01} \cdot {\bf v}_1 \right)
    \left( v_0^2 + v_1^2 \right)
\nonumber \\ 
&& \qquad
  + 22 v_0^2 v_1^2 
  + 4 {\left( {\bf v}_0\cdot{\bf v}_1 \right)}^2
  - 10 v_0^2 {\left( {\bf n}_{01}\cdot{\bf v}_1 \right)}^2
  - 10 v_1^2 {\left( {\bf n}_{01}\cdot{\bf v}_0 \right)}^2
\nonumber \\ 
&& \qquad
  + 24 \left( {\bf v}_0\cdot{\bf v}_1 \right)
    \left( {\bf n}_{01}\cdot{\bf v}_0 \right)
    \left( {\bf n}_{01}\cdot{\bf v}_1 \right)
  + 6 {\left( {\bf n}_{01}\cdot{\bf v}_0 \right)}^2
    {\left( {\bf n}_{01}\cdot{\bf v}_1 \right)}^2
\Big\}
\nonumber \\
&&
+
\frac{m_0 m_2}{16 r_{02}^2}
\Big\{
  14 \left( v_0^4 + v_2^4 \right)
  - 28 \left( {\bf v}_0\cdot{\bf v}_2 \right)
    \left( v_0^2 + v_2^2 \right)
  - 4 \left( {\bf n}_{02} \cdot {\bf v}_0 \right)
    \left( {\bf n}_{02} \cdot {\bf v}_2 \right)
    \left( v_0^2 + v_2^2 \right)
\nonumber \\ 
&& \qquad
  + 22 v_0^2 v_2^2 
  + 4 {\left( {\bf v}_0\cdot{\bf v}_2 \right)}^2
  - 10 v_0^2 {\left( {\bf n}_{02}\cdot{\bf v}_2 \right)}^2
  - 10 v_2^2 {\left( {\bf n}_{02}\cdot{\bf v}_0 \right)}^2
\nonumber \\ 
&& \qquad
  + 24 \left( {\bf v}_0\cdot{\bf v}_2 \right)
    \left( {\bf n}_{02}\cdot{\bf v}_0 \right)
    \left( {\bf n}_{02}\cdot{\bf v}_2 \right)
  + 6 {\left( {\bf n}_{02}\cdot{\bf v}_0 \right)}^2
    {\left( {\bf n}_{02}\cdot{\bf v}_2 \right)}^2
\Big\}\nonumber \\
&&
+
\frac{m_1 m_2}{16 r_{12}^2}
\Big\{
  14 \left( v_1^4 + v_2^4 \right)
  - 28 \left( {\bf v}_1\cdot{\bf v}_2 \right)
    \left( v_1^2 + v_2^2 \right)
  - 4 \left( {\bf n}_{12} \cdot {\bf v}_1 \right)
    \left( {\bf n}_{12} \cdot {\bf v}_2 \right)
    \left( v_1^2 + v_2^2 \right)
\nonumber \\ 
&& \qquad
  + 22 v_1^2 v_2^2 
  + 4 {\left( {\bf v}_1\cdot{\bf v}_2 \right)}^2
  - 10 v_1^2 {\left( {\bf n}_{12}\cdot{\bf v}_2 \right)}^2
  - 10 v_2^2 {\left( {\bf n}_{12}\cdot{\bf v}_1 \right)}^2
\nonumber \\ 
&& \qquad
  + 24 \left( {\bf v}_1\cdot{\bf v}_2 \right)
    \left( {\bf n}_{12}\cdot{\bf v}_1 \right)
    \left( {\bf n}_{12}\cdot{\bf v}_2 \right)
  + 6 {\left( {\bf n}_{12}\cdot{\bf v}_1 \right)}^2
    {\left( {\bf n}_{12}\cdot{\bf v}_2 \right)}^2
\Big\}\, .
\label{EQ:2PNL11}
\end{eqnarray}
%
%


\subsection{The transverse--traceless part of the Lagrangian $U_{TT}$}
\label{SEC:UTT}

The calculation of the 3-body ADM Lagrangian from \citet{DAMOUR85} is
somewhat lenghty, but straightforward, except for the
transverse-traceless part of the interaction potential $U_{TT}$ in
equations (\ref{EQ:NBODYL4}) and (\ref{EQ:2PNL04}). It is not known in
general how to evaluate this term explicitly, except for the two-body
case (see \citet{DAMOUR85,OHTA73,OHTA74}).  In order to circumvent
this problem, and since we are interested only in the vicinity of
the star with mass $m_1$, we assume
\begin{eqnarray}
\label{EQ:APPROXR}
  r_{01} \ll r_{12}\, , &\textrm{and}& r_{01} \ll r_{02}\, .
\end{eqnarray}
and expand equations (\ref{EQ:NBODYL2}) and (\ref{EQ:NBODYL4}) in
terms of $r_{01}/r_{12}$.
As in the rest of the Lagrangian, we assume the body
0 to be a point-particle and therfore drop all terms that are of
quadratic (or higher) power in $m_0$.

Utilizing these two physically motivated assumptions, we expand
$U_{TT}$ as
\begin{eqnarray}
U_{TT} = U_{TT}^{(12,12)} + U_{TT}^{(10,12)} + U_{TT}^{(12,02)} +
{\mathcal O}(m_0^2)\, ,
\label{EQ:UTT}
\end{eqnarray}
with
\begin{eqnarray}
U_{TT}^{(AB,CD)}&=& -\frac{1}{4\pi}
\int{d^3x
{\left[f_{ij}^{TT}(A,B)\right]}_{,k}
{\left[f_{ij}^{TT}(C,D)\right]}_{,k}
}
\nonumber \\
&=& \frac{1}{2\pi}
\int{d^3x
{\left(\frac{m_A}{r_A}\right)}_{,i}
{\left(\frac{m_B}{r_B}\right)}_{,j}
f_{ij}^{TT}(C,D)
}\, ,
\end{eqnarray}
where $A, B, C, D = 0, 1, 2$ and $g_{,i}\equiv \partial g(x)/\partial
x^i$ for any function $g(x)$.

The integrand $f_{ij}^{TT}(A,B)$ is given by
\begin{eqnarray}
f_{ij}^{TT}(A,B)&=&
\left(
\frac{\partial}{\partial R_A^i} \frac{\partial}{\partial R_B^j} 
+
\frac{\partial}{\partial R_A^j} \frac{\partial}{\partial R_B^i} 
\right)
\ln \left(r_A + r_B + r_{AB} \right) 
\nonumber \\
&&
- \frac{\delta_{ij}}{8} 
\left(
\frac{2}{r_A r_B} - \frac{2}{r_A r_{AB}} - 
\frac{2}{r_B r_{AB}} 
+ 2\frac{{\bf n}_A\cdot {\bf n}_{AB} }{r_{AB}^2} - 
2\frac{{\bf n}_B\cdot {\bf n}_{AB} }{r_{AB}^2} 
\right)
\nonumber \\
&&
- \frac{\partial_{ij}^2}{8} 
\left(
2\, \ln \left(r_A + r_B + r_{AB} \right) -
\frac{r_A + r_B}{r_{AB}} 
+ \frac{r_A^2 ({\bf n}_A\cdot {\bf n}_{AB}) }{2r_{AB}^2}  
- \frac{r_B^2 ({\bf n}_B\cdot {\bf n}_{AB}) }{2r_{AB}^2}  
\right)
\nonumber \\
&&
+ \frac{1}{2r_{AB}^2}
\left(
n_{A}^i\, n_{AB}^j + n_{A}^j\, n_{AB}^i
- n_{B}^i\, n_{AB}^j - n_{B}^j\, n_{AB}^i
\right)\, ,
\end{eqnarray}
where $i=1, 2, 3$ denotes spatial components. Here, as in Figure
\ref{FIG:3BODY},
\begin{eqnarray}
{\bf r}_A={\bf r} - {\bf R}_A\, ,
\quad
{\bf R}_A={\bf R}_B + {\bf R}_{AB}\, ,
\quad
\textrm{and}
\quad
{\bf r}_A={\bf r}_B + {\bf R}_{AB}.
\label{EQ:VECTORS}
\end{eqnarray}
In the preceding equations, we denote the position of the integration
point with ${\bf r}$, the position (i.e. ``trajectory'') of the body
$A$ with ${\bf R}_A$, the vector between the integration point and the
body $A$ with ${\bf r}_A$, and the vector defined by two bodies $A$
and $B$ with ${\bf R}_{AB}$.
 
Two out of three terms in equation (\ref{EQ:UTT}) are already
available or easily derived from previous
calculations of the
two-body Lagrangian. The result for $U_{TT}^{(12,12)}$ is given in
\citet{DAMOUR85,OHTA73,OHTA74}:
\begin{eqnarray}
U_{TT}^{(12,12)} 
&=&
-\frac{1}{2}\, \frac{m_1^2 m_2^2}{r_{12}^3}\, .
\end{eqnarray}
It is straightforward to obtain an approximate result for
$U_{TT}^{(12,02)}$ by using an expansion in terms of $r_{\sss
01}/r_{\sss 12}$ and keeping only terms with nonpositive powers of
$r_{\sss 01}/r_{\sss 12}$.  This approximation has a straightforward
physical interpretation that $r_{\sss 02}\approx r_{\sss 12}$ around
${\bf R}_1$. A quick calculation yields
\begin{eqnarray}
U_{TT}^{(12,02)}
&\approx&
-\frac{1}{2}\, \frac{m_0 m_1 m_2^2}{r_{12}^3}
+{\mathcal O}\left(
\frac{r_{01}}{r_{12}}
\right)\, .
\end{eqnarray}
However, the remaining term turns out to require a much more laborious
calculation. We start from the expression
\begin{eqnarray}
U_{TT}^{(10,12)} 
&=& 
2
\int{
\frac{d^3x}{4\pi} 
\left( \frac{m_2}{r_2} \right)_{,i} 
\left( \frac{m_1}{r_1} \right)_{,j}
f_{ij}^{TT}(1, 0) 
}
\nonumber \\
&=& 
2 m_0 m_1^2 m_2 
\int{
\frac{d^3x}{4\pi} 
\frac{x^i - R_2^i}{r_2^3}\,
\frac{x^j - R_1^j}{r_1^3}
} 
\nonumber \\
&&\times \, 
\left\{
\left(
\frac{\partial}{\partial R_1^i} \frac{\partial}{\partial R_0^j} 
+
\frac{\partial}{\partial R_1^j} \frac{\partial}{\partial R_0^i} 
\right)
\ln \left(r_1 + r_0 + r_{01} \right) \right.
\nonumber \\
&& \quad \left.
- \frac{\delta_{ij}}{8} 
\left(
\frac{2}{r_1 r_0} - \frac{2}{r_1 r_{01}} - 
\frac{2}{r_0 r_{01}} 
+ 2\frac{{\bf n}_0\cdot {\bf n}_{01} }{r_{01}^2} - 
2\frac{{\bf n}_1\cdot {\bf n}_{01} }{r_{01}^2} 
\right)
\right.
\nonumber \\
&&\quad \left.
- \frac{\partial_{ij}^2}{8} 
\left(
2\, \ln \left(r_1 + r_0 + r_{01} \right) -
\frac{r_1 + r_0}{r_{01}} 
+ \frac{r_0^2 ({\bf n}_0\cdot {\bf n}_{01}) }{2r_{01}^2}  
- \frac{r_1^2 ({\bf n}_1\cdot {\bf n}_{01}) }{2r_{01}^2}  
\right)
\right.
\nonumber \\
&&\quad \left.
+ \frac{1}{2r_{01}^2}
\left(
n_{0}^i\, n_{01}^j + n_{0}^j\, n_{01}^i
- n_{1}^i\, n_{01}^j - n_{1}^j\, n_{01}^i
\right)
\right\}\, .
\label{EQ:UTTHARDTERM}
\end{eqnarray}
Instead of performing the entire calculation, we present
all the necessary ingredients and techniques for an (alas, still
long!) example.  As an illustration, we show details of the
calculation for the logarithmic term in equation
(\ref{EQ:UTTHARDTERM}). The procedures for the remaining terms are
identical and are omitted here for brevity.

After differentiating the logarithmic term with respect to body
positions ${\bf R}_A$, expanding the result in terms of $r_{\sss
01}/r_{\sss 12}$, and grouping terms according to their composition in
terms of $n_{\sss AB}^i$, we obtain
\begin{eqnarray}
&&2\int{\frac{d^3x}{4\pi}} 
\frac{n_1^i}{r_1^2}\,
\frac{n_2^j}{r_2^2}
\left(
\frac{\partial}{\partial R_1^i} \frac{\partial}{\partial R_0^j} 
+
\frac{\partial}{\partial R_1^j} \frac{\partial}{\partial R_0^i} 
\right)
\ln \left(r_1 + r_0 + r_{01} \right)
\nonumber \\
&&\qquad= 
2\int{
\frac{d^3x}{4\pi} 
\frac{n_1^i}{r_1^2}\,
\frac{n_2^j}{r_2^2}
\Bigg[
\frac{(n_1^i+n_{01}^i)(-n_0^j+n_{01}^j)}
{{(r_1+r_0+r_{01})}^2}}
+\frac{1}{r_1+r_0+r_{01}}\,
\frac{n_{01}^in_{01}^j - \delta^{ij}}
{r_{01}} \quad + \quad \Big(i\leftrightarrow j\Big)
\Bigg]
\nonumber \\
&&
\qquad
\approx 
2\int{}
\frac{d^3x}{4\pi} 
\frac{n_1^i}{r_1^2}\,
\frac{n_2^j}{r_2^2}
\Bigg[
-\frac{1}{4}
\frac{(n_1^i+n_{01}^i)(n_1^j-n_{01}^j)}{r_1^2}
+
\frac{1}{2}
\frac{n_{01}^i n_{01}^j - \delta^{ij}}{r_{01}r_1}
\quad + \quad \Big(i\leftrightarrow j\Big)
\Bigg]
\nonumber \\
&&\qquad= 
\frac{2}{r_{01}}
\big(n_{01}^i n_{01}^j - \delta^{ij}\big)
\int{
\frac{d^3x}{4\pi} 
\frac{n_1^i}{r_1^3}\,
\frac{n_2^j}{r_2^2}}
+
\big(n_{01}^i n_{01}^j - \delta^{ij}\big)
n_{01}^k
\int{
\frac{d^3x}{4\pi} 
\frac{n_1^in_1^k}{r_1^4}\,
\frac{n_2^j}{r_2^2}
}\, ,
\label{EQ:UTTLOGTERM}
\end{eqnarray}
where terms ${\mathcal O}(r_{\sss 01}/r_{\sss 12})$ have been dropped.

Various combinations of unit vectors can be expressed through
derivatives with respect to particles' distances about the integration
point $P$ 
\begin{eqnarray}
\frac{n_A^i}{r_A^N}
&=&
\frac{1}{N-1}
\tilde{\partial}_i
\left(\frac{1}{r_A^{N-1}}\right)\, ,
\label{EQ:UNITVECTOR1}
\end{eqnarray}
where $\tilde{\partial}_i\equiv \partial/\partial R_A^i$.  For
combinations of several $n_A^i$'s, analogous relations can be derived
to be:
\begin{eqnarray}
\frac{n_A^i n_A^j}{r_A^N}
&=&
\frac{1}{N(N-2)}
\tilde{\partial}^2_{ij}
\left(\frac{1}{r_A^{N-2}}\right)
+
\frac{1}{N}
\frac{\delta^{ij}}{r_A^N}\, ,
\label{EQ:UNITVECTOR2}
\\
\frac{n_A^i n_A^j n_A^k}{r_A^N}
&=&
\frac{1}{(N-3)(N-1)(N+1)}
\tilde{\partial}^3_{ijk}
\left(\frac{1}{r_A^{N-3}}\right)
\nonumber \\
&& \qquad
+
\frac{1}{(N-1)(N+1)}
\left(
\delta^{ij}\tilde{\partial}_k
+\delta^{jk}\tilde{\partial}_i
+\delta^{ki}\tilde{\partial}_j
\right)
\left(\frac{1}{r_A^{N-1}}\right)\, ,  
\label{EQ:UNITVECTOR3}
\\
\frac{n_A^i n_A^j n_A^k n_A^l}{r_A^N}
&=&
\frac{1}{(N-4)(N-2)N(N+2)}
\tilde{\partial}^4_{ijkl}
\left(\frac{1}{r_A^{N-4}}\right)
\nonumber \\ &&
+
\frac{1}{(N-2)N(N+2)}
\left(
\delta^{ij}\tilde{\partial}^2_{kl}+
\delta^{li}\tilde{\partial}^2_{jk}+
\delta^{kl}\tilde{\partial}^2_{ij}
\right.
\nonumber \\
&& \qquad
+
\left.
\delta^{jk}\tilde{\partial}^2_{li}+
\delta^{ik}\tilde{\partial}^2_{jl}+
\delta^{jl}\tilde{\partial}^2_{ik}
\right)
\left(\frac{1}{r_A^{N-2}}\right)
\nonumber \\ &&
+
\frac{1}{N(N-2)}\,
\frac{1}{r_A^N}\,
\left(
\delta^{ij}\delta^{kl}+
\delta^{ik}\delta^{jl}+
\delta^{il}\delta^{kj}
\right) \label{EQ:UNITVECTOR4} \, ,
\end{eqnarray}
where in equations (\ref{EQ:UNITVECTOR2}) through
(\ref{EQ:UNITVECTOR4}) we have abbreviated multiple derivatives as
\begin{eqnarray}
\tilde{\partial}_{ij\ldots k}^N
&\equiv&
\underbrace{
\frac{\partial}{\partial R_A^i} 
\frac{\partial}{\partial R_A^j} \ldots
\frac{\partial}{\partial R_A^k} 
}_{N\,{\rm derivatives}}\, .
\end{eqnarray}

After expressing the unit vectors in equation (\ref{EQ:UTTLOGTERM}) as
derivatives and taking derivatives in front of the integral sign, all
integrations are performed over integrands that are combinations of
powers of $r_1$ and $r_2$ only. Such integrals can be evaluated by the
use of the formula (see \cite{DAMOUR85})
\begin{eqnarray}
I(\alpha,\beta)&=&
\int{\frac{d^3x}{4\pi} r_A^\alpha r_B^\beta 
}
= 
\frac{\sqrt{\pi}}{4}
\frac{
\Gamma\left(\frac{\alpha+3}{2}\right) 
\Gamma\left(\frac{\beta+3}{2}\right) 
\Gamma\left(-\frac{\alpha + \beta + 3}{2} \right)
} 
{
\Gamma\left(-\frac{\alpha}{2}\right) 
\Gamma\left(-\frac{\beta}{2}\right) 
\Gamma\left(\frac{\alpha + \beta + 6}{2} \right)
}
r_{AB}^{\alpha+\beta+3}\, .
\label{EQ:GAMMAINTEGRAL}
\end{eqnarray}

After performing integrations, differentiations with respect to body
trajectories $R_A$ have to be performed. Integrations yield results
that depend on $R_A$ as powers of $r_{AB}^N={\left|{\bf R}_A - {\bf
R}_B\right|}^N$ and in order to calculate $U_{TT}^{(10,12)}$, we need
to perform up to five consecutive differentiations with respect to
$R_A$:
\begin{eqnarray}
\frac{\partial}{\partial R_A^i}\,\frac{1}{r_{AB}^N}
&=&
-N\frac{n_{AB}^i}{r_{AB}^{N+1}}\, ,
\label{EQ:DIFFR1}
\\
\frac{\partial}{\partial R_A^i}
\frac{\partial}{\partial R_A^j}
\,\frac{1}{r_{AB}^N}
&=&
\frac{N}{r_{AB}^{N+2}}
\left((N+2)n_{AB}^in_{AB}^j - \delta{ij} \right)\, ,
\label{EQ:DIFFR2}
\\
\frac{\partial}{\partial R_A^i}
\frac{\partial}{\partial R_A^j}
\frac{\partial}{\partial R_A^k}
\,\frac{1}{r_{AB}^N}
&=&
\frac{N(N+2)}{r_{AB}^{N+3}}
\left(
\delta^{ij}n_{AB}^k +
\delta^{ki}n_{AB}^j +
\delta^{jk}n_{AB}^i 
\right.
\nonumber \\
&& \qquad
\left.
-
(N+4)\,
n_{AB}^i n_{AB}^j n_{AB}^k
\right)\, ,
\label{EQ:DIFFR3}
\\
\frac{\partial}{\partial R_A^i}
\frac{\partial}{\partial R_A^j}
\frac{\partial}{\partial R_A^k}
\frac{\partial}{\partial R_A^l}
\,\frac{1}{r_{AB}^N}
&=&
\frac{N(N+2)}{r_{AB}^{N+4}}
\Bigg[
\delta^{ij} \delta^{kl} +
\delta^{ik} \delta^{jl} +
\delta^{il} \delta^{jk} 
+ 
(N+4)(N+6)\,
n_{AB}^{ijkl}
\nonumber \\ 
&& \qquad
- 
(N+4)\, \left(
\delta^{ij} n_{AB}^{kl} +
\delta^{ik} n_{AB}^{jl} +
\delta^{il} n_{AB}^{jk} +
\delta^{jk} n_{AB}^{il} 
\right.
\nonumber \\ 
&& \qquad
+
\left.
\delta^{jl} n_{AB}^{ik} +
\delta^{kl} n_{AB}^{ij} 
\right)
\Bigg]\, ,
\label{EQ:DIFFR4}
\\
\frac{\partial}{\partial R_A^i}
\frac{\partial}{\partial R_A^j}
\frac{\partial}{\partial R_A^k}
\frac{\partial}{\partial R_A^l}
\frac{\partial}{\partial R_A^m}
\,\frac{1}{r_{AB}^N}
&=&
-\frac{N(N+2)(N+4)}{r_{AB}^{N+5}}
\Bigg\{
\left(
\delta^{ij} \delta^{kl} +
\delta^{ik} \delta^{jl} +
\delta^{il} \delta^{jk} 
\right) n_{AB}^{m} 
\nonumber \\ 
&& \qquad
+ 
\left(
\delta^{mj} \delta^{kl} +
\delta^{mk} \delta^{jl} +
\delta^{ml} \delta^{jk} 
\right) n_{AB}^{i} 
\nonumber \\ 
&& \qquad
+ 
\left(
\delta^{im} \delta^{kl} +
\delta^{ik} \delta^{ml} +
\delta^{il} \delta^{mk} 
\right) n_{AB}^{j} 
\nonumber \\ 
&& \qquad
+ 
\left(
\delta^{ij} \delta^{ml} +
\delta^{im} \delta^{jl} +
\delta^{il} \delta^{jm} 
\right) n_{AB}^{k} 
\nonumber \\ 
&& \qquad
+ 
\left(
\delta^{ij} \delta^{km} +
\delta^{ik} \delta^{jm} +
\delta^{im} \delta^{jk} 
\right) n_{AB}^{l}
\nonumber \\ 
&& \qquad
- 
(N+6)\,
\Big[
\delta^{ij} n_{AB}^{klm} + \delta^{ik} n_{AB}^{jlm} +
\delta^{il} n_{AB}^{jkm} + \delta^{im} n_{AB}^{jkl} 
\nonumber \\ 
&& \qquad 
+
\delta^{jk} n_{AB}^{ilm} + \delta^{jl} n_{AB}^{ikm} + 
\delta^{jm} n_{AB}^{ikl} 
\nonumber \\ 
&& \qquad 
+ 
\delta^{kl} n_{AB}^{ijm} + \delta^{km} n_{AB}^{ijl} + 
\delta^{lm} n_{AB}^{ijk} 
\Big]
\nonumber \\ 
&& \qquad
+
(N+6)(N+8)\, n_{AB}^{ijklm} 
\Bigg\}\, ,
\label{EQ:DIFFR5}
\end{eqnarray}
where $n_{AB}^{ij\ldots m}\equiv n_{AB}^i n_{AB}^j\ldots n_{AB}^m$. We
also note that differentiations with respect to $R_1^i$ and $R_2^i$
can be related through the identity
\begin{eqnarray}
\frac{\partial}{\partial R_B^i}\,\frac{1}{r_{AB}^N}
&=&
-\frac{\partial}{\partial R_A^i}\,\frac{1}{r_{AB}^N}\, .
\end{eqnarray}

We now assemble these results. Firstly, after rewriting the
unit vectors as derivatives through equations (\ref{EQ:UNITVECTOR1})
and (\ref{EQ:UNITVECTOR2}), we perform the necessary integrations by
employing equation (\ref{EQ:GAMMAINTEGRAL}):
\begin{eqnarray}
&&
\frac{2}{r_{01}}
\big(n_{01}^i n_{01}^j - \delta^{ij}\big)
\int{
\frac{d^3x}{4\pi} 
\frac{n_1^i}{r_1^3}\,
\frac{n_2^j}{r_2^2}} 
=
\nonumber \\
&& \qquad
\lim_{\alpha,\beta\rightarrow 0}
\left[
\frac{2}{r_{01}}
\big(n_{01}^i n_{01}^j - \delta^{ij}\big)
\frac{1}{2-\alpha}
\frac{1}{1-\beta}
\frac{\partial}{\partial R_1^i}
\frac{\partial}{\partial R_2^j}
I(\alpha-2, \beta-1)
\right]\, ,
\end{eqnarray}
\begin{eqnarray}
&&
\big(n_{01}^i n_{01}^j - \delta^{ij}\big)
n_{01}^k
\int{
\frac{d^3x}{4\pi} 
\frac{n_1^in_1^k}{r_1^4}\,
\frac{n_2^j}{r_2^2}
} 
=
\nonumber \\ 
&& \qquad
\lim_{\alpha,\beta\rightarrow 0}
\Bigg[
\big(n_{01}^i n_{01}^j - \delta^{ij}\big)
n_{01}^k
\Bigg(
\frac{1}{4-\alpha}
\frac{1}{2-\alpha}
\frac{1}{1-\beta}
\frac{\partial}{\partial R_1^i}
\frac{\partial}{\partial R_2^j}
\frac{\partial}{\partial R_1^k}
I(\alpha-2, \beta-1)
\nonumber \\ 
&& \qquad
+
\frac{1}{4-\alpha} \delta^{ik}
\frac{\partial}{\partial R_2^j}
I(\alpha-4, \beta-1)
\Bigg)
\Bigg]\, .
\end{eqnarray}

Finally, the application of differentiations in equations
(\ref{EQ:DIFFR1}) through (\ref{EQ:DIFFR5}) yields the 
result
\begin{eqnarray}
&&
2\int{\frac{d^3x}{4\pi}} 
\frac{n_1^i}{r_1^2}\,
\frac{n_2^j}{r_2^2}
\left(
\frac{\partial}{\partial R_1^i} \frac{\partial}{\partial R_0^j} 
+
\frac{\partial}{\partial R_1^j} \frac{\partial}{\partial R_0^i} 
\right)
\ln \left(r_1 + r_0 + r_{01} \right)
\approx 
\nonumber \\ 
&& \qquad
-2\frac{{({\bf n}_{01}\cdot {\bf n}_{12})}^2}{r_{01}r_{12}^2}
-\frac{1}{2}\,\frac{{({\bf n}_{01}\cdot {\bf n}_{12})}}{r_{12}^3}
+\frac{{({\bf n}_{01}\cdot {\bf n}_{12})}^3}{r_{12}^3}\, .
\end{eqnarray}

In order perform the full calculation for $U_{TT}^{(10,12)}$, we need
to compute the remaining terms in equation
(\ref{EQ:UTTHARDTERM}). They can be calculated by applying the
technique described above on the following expressions:
\begin{eqnarray}
\partial_i \partial_j \ln \left(r_0 + r_1 + r_{01}\right) 
&\approx&
-2\frac{n_1^i n_1^j}{r_1^2} + \frac{\delta^{ij}}{r_1^2} \, ,
\\
\partial_i \partial_j \left(-\frac{r_0 + r_1}{r_{01}}\right) 
&\approx&
2\frac{n_1^i n_1^j - \delta^{ij}}{r_1 r_{01}} +
\frac{3 n_1^i n_1^j - \delta^{ij}}{r_1^2}({\bf n}_1\cdot{\bf n}_{01}) 
\nonumber \\ 
&& \qquad
-
\frac{n_1^i n_{01}^j}{r_1^2} -
\frac{n_1^j n_{01}^i}{r_1^2} \, ,
\\
\partial_i \partial_j 
\left(\frac{r_1({\bf r}_1\cdot{\bf r}_{01})}{2r_{01}^3}-
\frac{r_0({\bf r}_0\cdot{\bf r}_{01})}{2r_{01}^3}
\right) 
&\approx&
\frac{1}{2 r_1 r_{01}}
\Bigg\{
-2\left( n_1^i n_{01}^j + n_1^j n_{01}^i \right)
({\bf n}_1\cdot{\bf n}_{01})
\nonumber \\ 
&& \qquad
+
\delta^{ij} \left(1-{({\bf n}_1\cdot{\bf n}_{01})}^2\right)
\nonumber \\ &&\qquad
+
2 n_{01}^i n_{01}^j +
n_1^i n_1^j \left(3{({\bf n}_1\cdot{\bf n}_{01})}^2 - 1\right)
\Bigg\}
\nonumber \\ 
&& \qquad
+
\frac{1}{2 r_1^2}
\Bigg\{
\frac{3}{2}
\left( n_1^i n_{01}^j + n_1^j n_{01}^i \right)
\left(1-3{({\bf n}_1\cdot{\bf n}_{01})}^2\right)
\nonumber \\ 
&& \qquad
+
3 n_{01}^i n_{01}^j ({\bf n}_1\cdot{\bf n}_{01})
\nonumber \\ &&\qquad
+
\frac{3}{2}
\delta^{ij} 
({\bf n}_1\cdot{\bf n}_{01})
\left(1-{({\bf n}_1\cdot{\bf n}_{01})}^2\right)
\nonumber \\ &&\qquad
+
\frac{3}{2}
n_1^i n_1^j 
({\bf n}_1\cdot{\bf n}_{01})
\left(5{({\bf n}_1\cdot{\bf n}_{01})}^2 - 3\right)
\Bigg\}
\\
\frac{n_0^i n_{01}^j}{r_{01}^2}
&\approx&
\frac{n_1^i n_{01}^j}{r_{01}^2}
+
\frac{n_1^i n_{01}^j}{r_1 r_{01}}
({\bf n}_1\cdot{\bf n}_{01})
+
\frac{n_1^i n_{01}^j}{2 r_1^2}
\left(3{({\bf n}_1\cdot{\bf n}_{01})}^2 - 1\right)
\nonumber \\ 
&& \qquad
-
\frac{n_{01}^i n_{01}^j}{r_{01} r_1}
-
\frac{n_{01}^i n_{01}^j}{r_1^2}
({\bf n}_1\cdot{\bf n}_{01}) \, .
\end{eqnarray}

A straightforward, but somewhat lengthy, calculation yields the
following approximate result
\begin{eqnarray}
&2&\int{
\frac{d^3x}{4\pi} 
\frac{n_1^i}{r_1^2}\,
\frac{n_2^j}{r_2^2}
} 
\Big\{
- \frac{\delta_{ij}}{8} 
\left(
\frac{2}{r_1 r_0} - \frac{2}{r_1 r_{01}} - 
\frac{2}{r_0 r_{01}} 
+ 2\frac{{\bf n}_0\cdot {\bf n}_{01} }{r_{01}^2} - 
2\frac{{\bf n}_1\cdot {\bf n}_{01} }{r_{01}^2} 
\right)
\nonumber \\
&&
- \frac{\partial_{ij}^2}{8} 
\left(
2\, \ln \left(r_1 + r_0 + r_{01} \right) -
\frac{r_1 + r_0}{r_{01}} 
+ \frac{r_0^2 ({\bf n}_0\cdot {\bf n}_{01}) }{2r_{01}^2}  
- \frac{r_1^2 ({\bf n}_1\cdot {\bf n}_{01}) }{2r_{01}^2}  
\right)
\nonumber \\
&& 
+ \frac{1}{2r_{01}^2}
\left(
n_{0}^i\, n_{01}^j + n_{0}^j\, n_{01}^i
- n_{1}^i\, n_{01}^j - n_{1}^j\, n_{01}^i
\right)
\Big\}
\nonumber \\ && \qquad
\approx
\frac{11}{16}\,\frac{1}{r_{01}r_{12}^2}
-
\frac{{({\bf n}_{01}\cdot{\bf n}_{12})}^2}{r_{01}r_{12}^2}
-
\frac{9}{32}\,
\frac{({\bf n}_{01}\cdot{\bf n}_{12})}{r_{12}^3}
-
\frac{7}{48}\,
\frac{{({\bf n}_{01}\cdot{\bf n}_{12})}^3}{r_{12}^3}\, .
\end{eqnarray}
The final result for the transverse--traceless term is
\begin{eqnarray}
U_{TT} 
&\approx&
U_{TT}^{(12,12)} + U_{TT}^{(10,12)} + U_{TT}^{(12,02)}
\nonumber \\
&\approx&
-\frac{1}{2}\, \frac{m_1^2 m_2^2}{r_{12}^3}
-\frac{1}{2}\, \frac{m_0 m_1 m_2^2}{r_{12}^3}
\nonumber \\
&&+\,
m_0 m_1^2 m_2
\Big[
\frac{11}{16}\,\frac{1}{r_{01}r_{12}^2}
-
3\frac{{({\bf n}_{01}\cdot{\bf n}_{12})}^2}{r_{01}r_{12}^2}
-
\frac{25}{32}\,
\frac{({\bf n}_{01}\cdot{\bf n}_{12})}{r_{12}^3}
+
\frac{41}{48}\,
\frac{{({\bf n}_{01}\cdot{\bf n}_{12})}^3}{r_{12}^3}
\Big] \, .
\label{EQ:UTTFINAL}
\end{eqnarray}


\section{Roche Lobes for 2PN}
\label{SEC:2PNROCHE}

\begin{figure}[ht!]
\begin{center}
\includegraphics[width=0.9\textwidth]{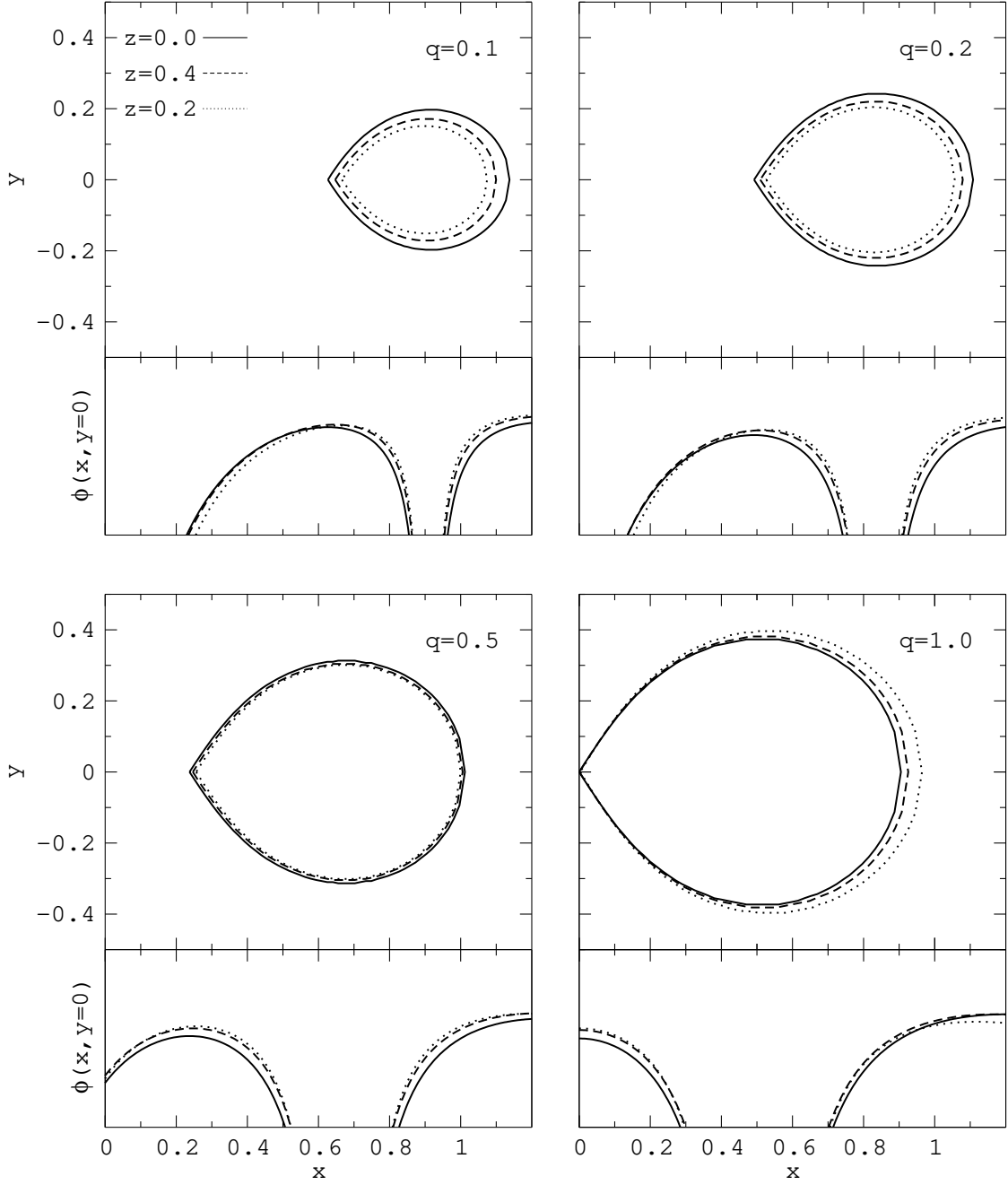}
\caption{Roche lobes and the corresponding potentials for
$y=0$. Coordinates $x$ and $y$ are scaled by the stellar separation
$a$ and are shown for $q = 0.1$, $0.2$, $0.5$, and $1.0$,
respectively. Results shown are for values of $z=0$, $z = 0.2$, and
$z=0.4$, respectively.
\label{FIG:EQUIP2PN}}
\end{center}
\end{figure}

We are now ready to find the equipotential surfaces that correspond to
the Roche lobes and to compute volumes that are contained inside the
lobes. We employ equations (\ref{EQ:2PNL01}) through
(\ref{EQ:2PNL11}), supplemented with equation
(\ref{EQ:UTTFINAL}). After setting the three bodies on quasicircular
orbits around the center of mass, we obtain the potential that gives
rise to forces that act on the point--particle of mass $m_0$. In
Figure \ref{FIG:EQUIP2PN}, we show the potential for $y=0$ (we
position the three bodies on the $x$--$y$ plane so $z=0$
automatically) and the corresponding Roche lobes for mass ratios
$q=0.1$, $0.2$, $0.5$, and $1.0$, respectively. Cusps on equipotential
lines correspond to the first Lagrange points $L_1$.  All distances
are scaled by the stellar radial separation $a\equiv r_{12}$. Note
that we only show the lobe around the first star, since our
approximation ($r_{01}\ll r_{12}$) is valid only in this region.

As in the Newtonian case, the volume within the equipotential surface
(Roche volume) grows with $q$ for fixed $a$. However, the potential
and the equipotential surfaces acquire an additional
dependence. Unlike for the Newtonian case, the total mass modifies the
result: for low $q$, the Roche volumes become smaller as the total
mass increases, whereas for $q$ greater than about 0.7 the volumes
increase. As for coordinate positions, where we have eliminated the
separation $a$, we can introduce a new dimensionless
parameter 
\begin{eqnarray}
z
&\equiv &
\frac{2M}{a}\, ,
\end{eqnarray}
involving the ratio of the total mass $M$ and the separation 
$a$. This parameter also corresponds to the ratio of Schwarzschild
radius for $M$ and the separation distance $a$. 

\begin{figure}[ht!]
\begin{center}
\includegraphics[width=0.9\textwidth]{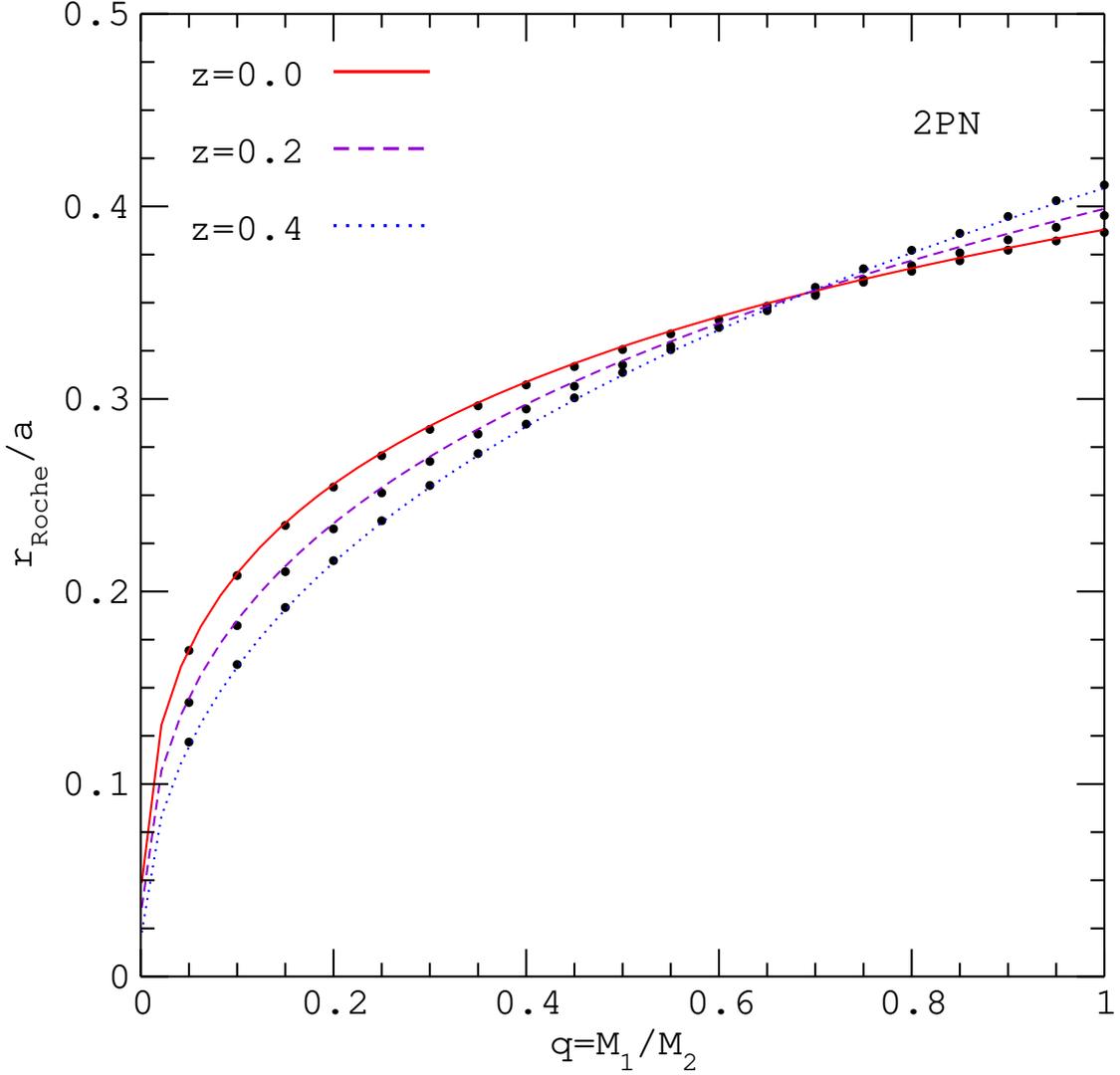}
\caption{Effective Roche lobe radii $r_{\sss Roche}$ scaled by the
stellar separation $a$ versus $q$ for $z=0$, $z = 0.2$, and $z=0.4$,
respectively.
\label{FIG:ROCHE2PN}}
\end{center}
\end{figure}

Integration of volumes enclosed by equipotential surfaces is
straightforward; we utilize a Newton--Cotes type algorithm to find the
enclosed volumes. Our results are shown in Figure \ref{FIG:ROCHE2PN}
for $q\in [0,1]$, and $z=0$, $z=0.2$, and $z=0.4$, respectively.

In Figure \ref{FIG:ROCHE2PN}, we also show results for the fitting
function. Following \citet{EGGLETON1}, we choose the parametrization
in which the scaled effective Roche radius is
\begin{eqnarray}\label{eq:roche2pn}
  r_{\sss Roche}/a & = & Q(q)\, C(q, z)\, ,
\end{eqnarray}
where
\begin{eqnarray}\label{eq:q2pn}
  Q(q) &=& \frac{\alpha_{\sss Q}\, q^{2/3} } {\beta_{\sss Q}\, q^{2/3} + \ln(1 + q^{1/3}
  )}\,  \label{EQ:Q2PN} 
\end{eqnarray}
is the fitting function previously given by \citet{EGGLETON1} and
\begin{eqnarray}
 C(q, z) & = & 1 + z\,(\alpha_{\sss C}\, q^{1/5} - \beta_{\sss C}) \, \label{EQ:C2PN}
\end{eqnarray}
is the correction function that stems from post-Newtonian effects.
Fitting parameters of $Q(q)$ are identical to the ones obtained by
\citet{EGGLETON1} ($\alpha_{\sss Q}=0.49$ and $\beta_{\sss Q}=0.57$)
and values of
\begin{eqnarray}
\alpha_{\sss C} = 1.951\, ,\qquad &\textrm{and}\qquad& 
\beta_{\sss C}  = 1.812\, 
\label{EQ:BC2PN} 
\label{EQ:AC2PN}
\end{eqnarray}
refer to the fitting function for $C(q, z)$. This functional form
describes extremely well the dependence of $q$ for $q < 1.0$. We note
that the crossover of reduced versus enlarged Roche lobes with respect to
the Newtonian ($z=0$) case occurs at $q=(\beta_C/\alpha_C)^5\approx
0.69$, in which case the Roche radius is virtually $z$--independent.

\subsection{Lagrange points and the center of mass}
\label{SEC:LPCM}
While performing the computation of the effective Roche volumes and
radii, it is necessary to find the first Lagrange point $L_1$. In
addition, there are two more extrema of the potential along the
$x$--axis that correspond to the second and third Lagrange points,
$L_2$ and $L_3$, respectively. Moreover, it is necessary to find the
position of the center of mass of the system while setting the
particles onto quasicircular orbits. We briefly outline below how the
positions of these points depend on the mass ratio and how the results
get modified compared to the Newtonian case. We consider only the
first three out of five Lagrange points ($L_1$, $L_2$, and $L_3$)
as $L_4$ and $L_5$ become local extrema only for nonvanishing
velocity ${\bf v}^{rot}_0$.

\subsubsection{Lagrange points}

\begin{figure}[ht!]
\begin{center}
\includegraphics[width=0.9\textwidth]{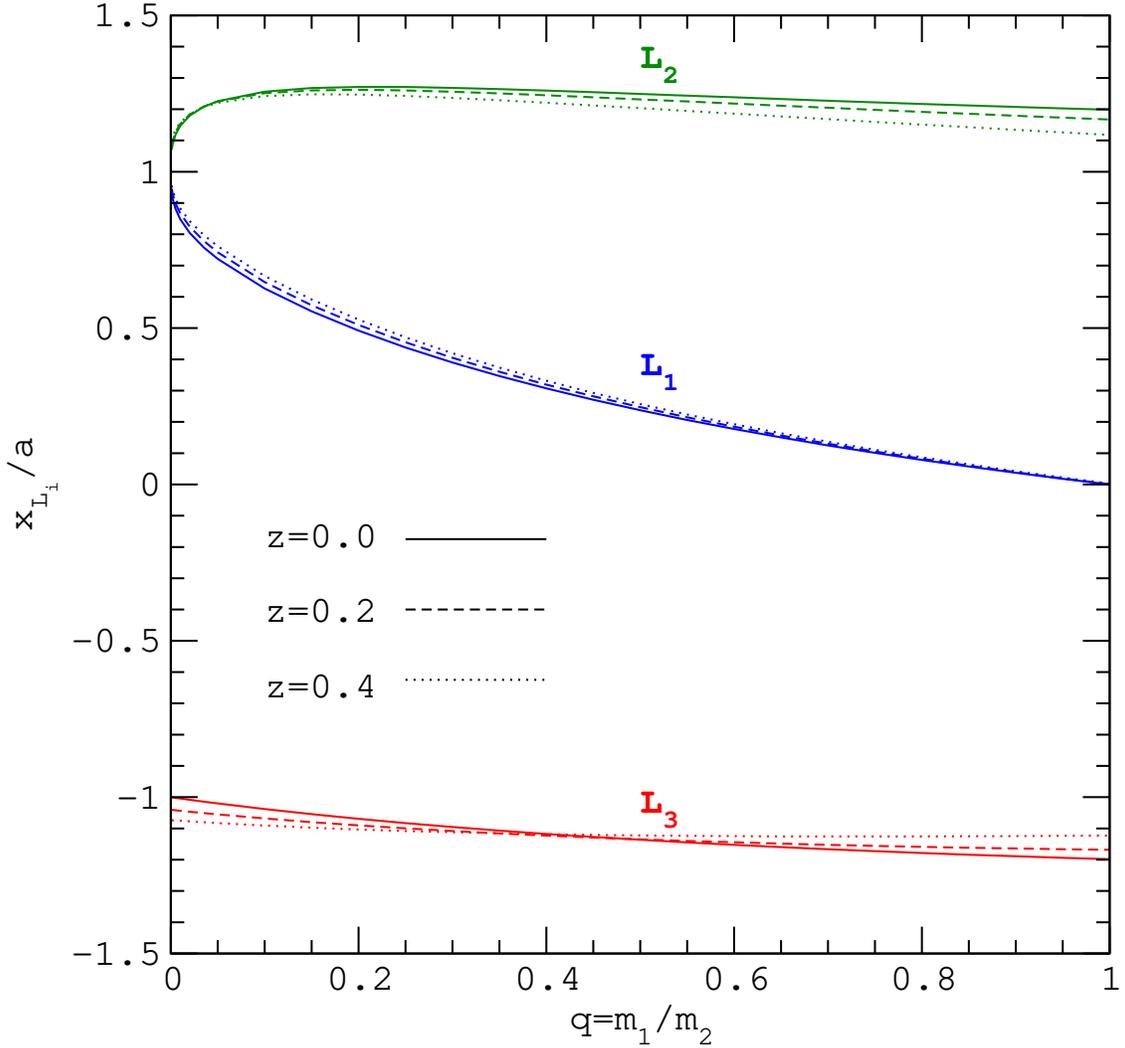}
\caption{The positions of the Lagrange points $L_1$, $L_2$, and $L_3$
as a function of the mass ratio $q=M_1/M_2$.  Results are for $z=0$,
$0.2$, and $0.4$, respectively.
\label{FIG:LAGRANGE}}
\end{center}
\end{figure}

In the Newtonian case, the positions of the Lagrange points can be
fully described in terms of the mass ratio $q=M_1/M_2$. As the mass
ratio drops towards $0$, the position of the first Lagrange point
quickly shifts away from the center of mass and in the direction of
the lighter star. The positions of Lagrange points $L_2$ and $L_3$
show a much weaker $q$--dependence as can be seen in Figure
\ref{FIG:LAGRANGE}, where the center of mass is positioned at
$x=0$. The second point $L_2$ shifts towards the lighter mass for
$q<0.2$ after it passes through the maximum of its distance from the
center of mass at $q\approx0.22$. The third Lagrange point is slowly,
but steadily, pulled toward the center of mass as $q\rightarrow 0$.

As we might expect from the Roche lobe analysis, effects of
post--Newtonian corrections give rise to an additional
$(M_1+M_2)/r_{12}=M/a$ dependence. In our model, this can be
parametrized with the quantity $z=2M/a$. In Figure \ref{FIG:LAGRANGE},
we show two additional sets of lines for $z=0.2$ and $0.4$ ($z=0$
corresponds to the Newtonian case). For all values of $q$ given in
Figure \ref{FIG:LAGRANGE}, the positions of $L_1$ and $L_2$ are
slightly pulled towards the lighter mass compared to the Newtonian
case. The position of $L_3$ is, however, pulled towards the center of
mass for $q\gtrsim 0.45$ and away from it for $q\lesssim 0.4$.

\subsubsection{Position of the center of mass}

In order to find the Roche lobes, we set the two stars on
quasicircular orbits and set the center of mass at the origin. The
post--Newtonian approximation modifies the position of the center of
mass compared to the Newtonian result since the conservation of total
momentum results in a slight shift of the position from the Newtonian
case. In this section, we show how important these corrections are and
parametrize them in terms of $z=2M/a$.

We denote the ratio of the distance between $m_1$ and the center of
mass by $a_1$. The separation between the two stars is $a$,
and their ratio is
\begin{eqnarray}
\frac{a_1}{a} = \beta\, , &\textrm{and}& \frac{a_2}{a} = 1 - \beta \, . 
\end{eqnarray}
Setting the center of mass at the origin yields the condition
\citep{BLANCHET2001}
\begin{eqnarray}
&&\beta m_1 
\Bigg\{
1+ \frac{{p_1}^2}{2{m_1}^2} - \frac{m_2}{2 a} -
\frac{{p_1}^4}{8{m_1}^4} + \frac{m_2}{4 a} \left( -5
\frac{{p_1}^2}{{m_1}^2} - \frac{{p_2}^2}{{m_2}^2} + 7\frac{{\bf
p_1}\cdot{\bf p_2}}{m_1 m_2} \right) + \frac{{m_2}\left( m_1 + m_2
\right) }{a^2}
\Bigg\}
= \nonumber \\
&&(1-\beta) m_2
\Bigg\{
1+ \frac{{p_2}^2}{2{m_2}^2} - \frac{m_1}{2a} -
\frac{{p_2}^4}{8{m_2}^4} + \frac{m_1}{4 a} \left( -5
\frac{{p_2}^2}{{m_2}^2} - \frac {{p_1}^2}{{m_1}^2} + 7 \frac{{\bf
p_1}\cdot{\bf p_2}}{m_1 m_2} \right) + \frac{m_1(m_1 + m_2)}{a^2}
\Bigg\}\, ,
\label{EQ:2PNCM}
\nonumber \\
\end{eqnarray}
which contains an implicit dependence on $\beta$, since for
quasicircular motion we can write the velocities of the two stars as
\begin{eqnarray}
{v_1}^2={\beta}^{2}{a}^{2}{w}^{2} \, , &\textrm{and}&
{v_2}^2={(1-\beta)}^{2}{a}^{2}{w}^{2} \, ,
\end{eqnarray}
and their product as
\begin{eqnarray}
{\bf v_1}\cdot{\bf v_2}&=&-{\beta}(1-\beta){a}^{2}{w}^{2} \, ,
\end{eqnarray}
where at the 2PN level \citep{BLANCHET2003}
\begin{eqnarray}
\omega^2 
&=&
\frac{M}{a^3}
\left(
1 + 
\frac{M}{a}(\nu-3) + 
\left(\frac{M}{a}\right)^2
(\frac{21}{4}-\frac{5}{8}\nu+\nu^2)
\right)\, ,
\end{eqnarray}
with $\nu = q/(1+q)^2$. Squares of the two momenta
are
\begin{eqnarray}
{p_1}^2 &=& 
{{m_1}}^{2}{\beta}^{4}{a}^{4}\omega^{4}
+
\frac{1}{2} {{m_1}}^{2}a\omega^{2}\left( 2 a+5 {m_2} \right) {\beta}^{2}
+
\frac{7}{2} {{m_1}}^{2}a\omega^{2}{m_2}
\beta \, , \\
{p_2}^2 &=& 
{{m_2}}^{2}{a}^{4}\omega^{4}{\beta}^{4}-4
{{m_2}}^{2}{a}^{4}\omega^{4}{\beta}^{3}+ \left( 3
{{m_2}}^{2}{a}^{4}\omega^{4}+\frac{1}{2} {{m_2}}^{2}a\omega^{2} \left( 2 a+6
{a}^{3}\omega^{2}+5 {m_1} \right) \right) {\beta}^{2}
\nonumber \\
&& + \left( -\frac{1}{2} {{m_2}}^{2}a\omega^{2} \left( 2 a+6
{a}^{3}\omega^{2}+5 {m_1} \right) + \frac{1}{2} {{m_2}}^{2}a\omega^{2} \left(
-2 a-2 {a}^{3}\omega^{2}-12 {m_1} \right) \right) \beta
\nonumber \\
&& - \frac{1}{2} {{m_2}}^{2}a\omega^{2} \left( -2 a-2 {a}^{3}\omega^{2}-12
{m_1} \right)  \, .
\end{eqnarray}
The product of the two momenta can be computed to be
\begin{eqnarray}
{\bf p_1}\cdot{\bf p_2} &=&
{m_1} {m_2} {a}^{4}\omega^{4}{\beta}^{4}-2 {m_1} {m_2}
{\beta}^{3}{a}^{4}\omega^{4}+\frac{1}{4} {m_1} {m_2} a\omega^{2} \left( 6
{a}^{3}\omega^{2}+4 a+5 {m_1}+5 {m_2} \right) {\beta}^{2} \nonumber \\
&&+\frac{1}{4} {m_1}
{m_2} a\omega^{2} \left( -4 a-2 {a}^{3}\omega^{2}+2 {m_2}-12 {m_1}
\right) \beta-\frac{7}{4} {m_1} {{m_2}}^{2}a\omega^{2} \, ,
\end{eqnarray}
and their fourth powers (up to 2PN order) are
\begin{eqnarray}
{p_1}^4 &=& {m_1}^4 {v_1}^4 \, , \\
{p_2}^4 &=& {m_2}^4 {v_2}^4 \, .
\end{eqnarray}

Inserting these results into equation (\ref{EQ:2PNCM}), the position
of the center of mass can be obtained by solving the $5^{th}$ order
polynomial
\begin{eqnarray}
0 &=& a_0 + a_1\beta+ a_2\beta^2 + a_3\beta^3 + a_4\beta^4 +
a_5\beta^5 \, ,
\end{eqnarray}
with coefficients $a_i$ given by
\begin{eqnarray}
a_0 &=&
-16 {m_1}^2 m_2 -16 {m_2}^2 m_1 - 8 m_2 a^4 \omega^2 + 49 m_1 {m_2}^2
a^2 \omega^2 - 6 m_2 a^6 \omega^4 - 16 m_2 a^2 \nonumber \\
&& + 8 m_2 m_1 a - 28 m_2 a^3 \omega^2 m_1 + 120 {m_1}^2 m_2 a^2
\omega^2 + 20 m_1 m_2 a^{5} \omega^4 \, , \\
a_1 &=&
32 {m_2}^2 m_1 + 16 m_2 a^2 + 16 m_1 a^2+24 m_2 a^4 \omega^2 + 30 m_2
a^6 \omega^4 + 80 m_2 a^3 \omega^2 m_1 \nonumber \\
&&- 90 m_1 m_2 a^5 \omega^4 + 32 {m_1}^2 m_2 - 98 m_1 {m_2}^2 a^2
\omega^2 - 16 m_2 m_1 a - 230 {m_1}^2 m_2 a^2 \omega^2 \, , \\
a_2 &=&
51 {m_1}^2 m_2 a^2 \omega^2 + 146 m_1 m_2 a^5 \omega^4 - 24 m_2 a^4
\omega^2 - 81 m_1 {m_2}^2 a^2 \omega^2 \nonumber \\
&&- 72 m_2 a^3 \omega^2 m_1 - 60 m_2 a^6 \omega^4 \, , \\
a_3 &=&
48 m_2 a^3 \omega^2 m_1 + 8 a^4 \omega^2 m_1 + 60 m_2 a^6 \omega^4 +
10 m_1 {m_2}^2 a^2 \omega^2 + 10 {m_1}^2 m_2 a^2 \omega^2 \nonumber \\
&& + 8 m_2 a^4 \omega^2 - 84 m_1 m_2 a^5 \omega^4 \, , \\
a_4 &=&
-30 m_2 a^6 \omega^4 - 20 m_1 m_2 a^5 \omega^4 \, , \\
a_5 &=&
6 m_2 a^6 \omega^4 + 8 m_1 m_2 a^5 \omega^4 + 6 a^6 \omega^4 m_1 \, .
\end{eqnarray}

\begin{figure}[ht!]
\begin{center}
\includegraphics[width=0.9\textwidth, angle=0]{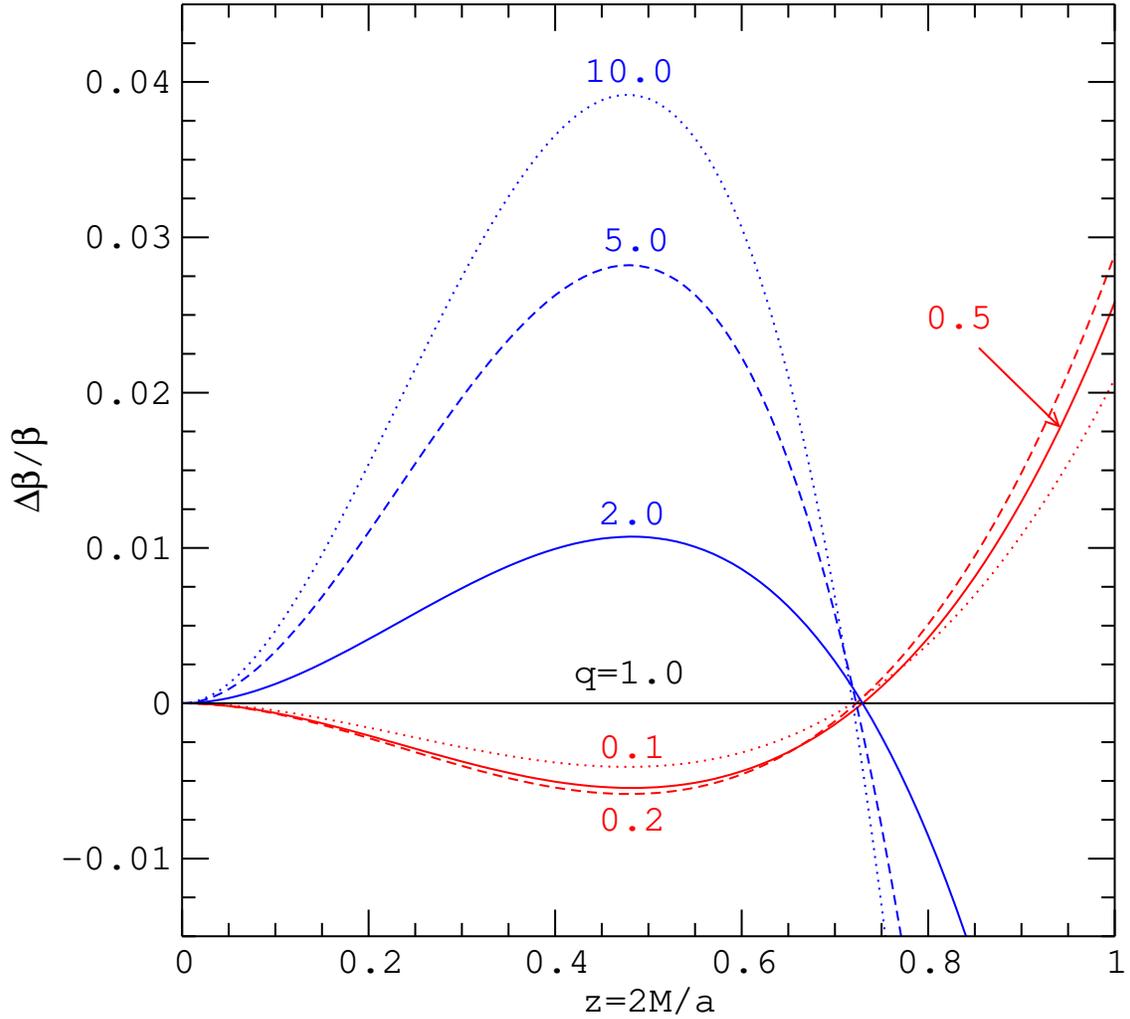}
\caption{ 
The position of the center of mass as a function of the relativity
parameter $z$ shown in the form of the ratio $\Delta\beta/\beta$ for
$q=0.1\ldots 10.0$. We note an increase of $\beta$ compared to the
Newtonian case for $q>1.0$. For $q>1.0$, $\beta$ decreases compared to
the Newtonian case.
\label{FIG:CM2PN}}
\end{center}
\end{figure}

The results obtained by numerical root-finding are shown in Figure
\ref{FIG:CM2PN}, in which we show the relative change of $\beta$
compared to the Newtonian case:
\begin{eqnarray}
\frac{\Delta \beta}{\beta}
&=&
\frac{\beta - \beta_N}{\beta}\, ,
\end{eqnarray}
where $\beta_N=m_2/M$. The dependence on the total mass $M$ and on the
separation $a$ is entirely through their ratio and is
parametrized in Figure \ref{FIG:CM2PN} through the parameter $z$.

The post--Newtonian analysis can be expected to be reliable up to
moderate values of $z$ ($z<0.4-0.5$). We note that in this regiont
post--Newtonian corrections to $\beta$ are less than about
$4\%$. Moreover, in all cases of interest ($q<1.0$), corrections
to $\beta$ are smaller than $0.5\%$.  As expected, deviations from the
Newtonian case increase for close or massive configurations, whereas
for large separations or low masses the results converge toward the
Newtonian case. Whereas for the more massive star ($q>1.0$) the
distance to the center of mass increases with $z$ compared to the
Newtonian case, the shift of the less massive star ($q<1.0$) does not
have a monotonic dependence on $q$. As expected from the symmetry of
the problem, corrections vanish for $q=1.0$.

\section{Conclusion}
\label{sec:conclusion}

We have utilized the second order post--Newtonian approximation in the
Arnowitt--Deser--Misner gauge to calculate Roche lobe volumes. These
results are an improvement over the Newtonian case in that
post--Newtonian gravity introduces corrections in the case of
moderately strong gravitational field.

In the course of our calculations, we have derived an approximate
three--body Lagrangian that is valid in the case when one of the
bodies is a point particle. This calculation requires an evaluation of
the transverse--traceless term $U_{TT}$ of the Lagrangian for which an
exact result is not available.  However, as shown in this work,
utilization of an approximation valid in the vicinity of the less
massive star enables this problem to be circumvented.

Using these results, we calculated Roche lobes in the 2PN effective potential
in the co--rotating frame and computed effective Roche lobe radii that can be
used to model mass transfer through Roche lobe overflow. In addition, we
computed changes to the positions of the Lagrange points and to the center of
mass due to post--Newtonian effects. We find that corrections to Newtonian
results for Roche lobe radii can be as significant as 20--30\% at low mass
ratio $q\lesssim 0.1$. Whereas for $q\gtrsim 0.7$ the Roche lobe radius
increases ($\approx 15\%$ for $q=1.0$), for low $q$'s the Roche lobe is
smaller than in the Newtonian case. We have provided our results in the form
of a simple fitting formula that depends on two physical parameters: the mass
ratio and the ratio of the total mass and the separation.

Research support of the U.S. Department of Energy under grant number
DOE/DE-FG02-87ER-40317 is gratefully acknowledged.

%
%



\bibliographystyle{apj} 

\bibliography{references}

\end{document}